\documentclass{aa}
\usepackage{graphics}
\usepackage{amssymb}
\newcommand{\kmprs}  {\mbox{\rm\,km\,s$^{-1}$}}
\newcommand{\feh} {\mbox{\rm [Fe/H]}}

\newcommand{\teff}  {\mbox{$T_{\rm eff}$}}
\newcommand{\logteff} {\mbox{${\rm log} T_{\rm eff}$}}
\newcommand{\logg}  {\mbox{{\rm log}$g$}}
\newcommand{\Ulsr} {\mbox{$U_{\rm LSR}$}}
\newcommand{\Vlsr} {\mbox{$V_{\rm LSR}$}}
\newcommand{\Wlsr} {\mbox{$W_{\rm LSR}$}}
\newcommand{\Rm} {\mbox{$R_{\rm m}$}}
\newcommand{\OFe} {\mbox{\rm [O/Fe]}}
\newcommand{\NaFe} {\mbox{\rm [Na/Fe]}}

\newcommand{\MgFe} {\mbox{\rm [Mg/Fe]}}
\newcommand{\AlFe} {\mbox{\rm [Al/Fe]}}
\newcommand{\KFe} {\mbox{\rm [K/Fe]}}
\newcommand{\SiFe} {\mbox{\rm [Si/Fe]}}
\newcommand{\CaFe} {\mbox{\rm [Ca/Fe]}}
\newcommand{\TiFe} {\mbox{\rm [Ti/Fe]}}

\newcommand{\NiFe} {\mbox{\rm [Ni/Fe]}}

\newcommand{\BaFe} {\mbox{\rm [Ba/Fe]}}

\newcommand{\ffe}  {\mbox{${\rm [\frac{Fe}{H}]}$}}
\newcommand{\ofe}  {\mbox{${\rm [\frac{O}{Fe}]}$}}
\newcommand{\nafe} {\mbox{${\rm [\frac{Na}{Fe}]}$}}
\newcommand{\mgfe} {\mbox{${\rm [\frac{Mg}{Fe}]}$}}
\newcommand{\alfe} {\mbox{${\rm [\frac{Al}{Fe}]}$}}
\newcommand{\sife} {\mbox{${\rm [\frac{Si}{Fe}]}$}}
\newcommand{\kfe} {\mbox{${\rm [\frac{K}{Fe}]}$}}
\newcommand{\vfe} {\mbox{${\rm [\frac{V}{Fe}]}$}}
\newcommand{\cafe} {\mbox{${\rm [\frac{Ca}{Fe}]}$}}
\newcommand{\tife} {\mbox{${\rm [\frac{Ti}{Fe}]}$}}
\newcommand{\crfe} {\mbox{${\rm [\frac{Cr}{Fe}]}$}}
\newcommand{\nife} {\mbox{${\rm [\frac{Ni}{Fe}]}$}}

\newcommand{\bafe} {\mbox{${\rm [\frac{Ba}{Fe}]}$}}

\newcommand{\Oone} {\ion{O}{i}}
\newcommand{\Naone} {\ion{Na}{i}}
\newcommand{\Mgone} {\ion{Mg}{i}}
\newcommand{\Alone} {\ion{Al}{i}}
\newcommand{\Sione} {\ion{Si}{i}}

\newcommand{\Kone} {\ion{K}{i}}
\newcommand{\Caone} {\ion{Ca}{i}}

\newcommand{\Tione} {\ion{Ti}{i}}

\newcommand{\Vone} {\ion{V}{i}}
\newcommand{\Crone} {\ion{Cr}{i}}

\newcommand{\Feone} {\ion{Fe}{i}}
\newcommand{\Fetwo} {\ion{Fe}{ii}}
\newcommand{\Nione} {\ion{Ni}{i}}

\newcommand{\Batwo} {\ion{Ba}{ii}}

\begin{document}
\thesaurus{07(02.14.1; 08.01.1; 08.11.1; 10.01.1; 10.05.1; 10.19.1)}
\title{Chemical composition of 90 F and G disk dwarfs 
\thanks{Based on observations carried out at the Beijing Astronomical
  Observatory (Xinglong, P.R. China). Tables 3, 4 and 5 are only
available in electronic form at the CDS via anonymous ftp to
cdsarc.u-strasbg.fr (130.79.128.5) or via http://cdsweb.u-strasbg.fr/Abstract.html}}
\author{Y.Q.\,Chen\inst{1} \inst{2} \inst{3} \and P.E.\,Nissen \inst{1} \and G.\,Zhao \inst{3} 
\and H.W.\,Zhang\inst{3} \inst{4} \and T.\,Benoni\inst{1}}
\offprints{P.E. Nissen (pen@obs.aau.dk)}
\institute{Institute of Physics and Astronomy, University of Aarhus, DK--8000
Aarhus C, Denmark
\and  Department of Astronomy, Beijing Normal University, Beijing 100875, P.R. China 
\and Beijing Astronomical Observatory, Chinese Academy of Sciences, Beijing 100012, P.R. China
\and Department of Geophysics, Peking University, Beijing 100871, P.R. China}
\date{Received October 7 1999 / Accepted November 22 1999}
\maketitle

\begin{abstract}
High resolution, high S/N spectra have been obtained for 
a sample of 90 F and G main-sequence disk stars covering the
metallicity range $-1.0 < \feh < +0.1$, and have been
analysed in a parallel way to the work of
Edvardsson et al. (\cite{Edvardsson93a}) in order to
re-inspect their results and to reveal new information on the
chemical evolution of the Galactic disk.

Compared to Edvardsson et al. the present study includes several improvements.
Effective temperatures are based on  the Alonso et al. (\cite{Alonso96})
calibration of color indices by the infrared flux method and surface 
gravities are calculated from Hipparcos parallaxes, which 
also allow more accurate ages to be calculated
from a comparison of $M_V$ and \teff\ with isochrones. In addition, more
reliable kinematical parameters are derived from Hipparcos distances and
proper motions in combination with accurate radial velocities. Finally,
a larger spectral coverage, 5600 - 8800\AA , makes it possible to
improve the abundance accuracy by studying more lines and to discuss
several elements not included in the work of Edvardsson et al.

The present paper provides the data and discusses some general results
of the abundance survey. A group of stars in the metallicity
range of $-1.0 < \feh < -0.6$ having
a small mean Galactocentric distance in the stellar orbits, $\Rm < 7$~kpc,
are shown to be older than the other disk stars and probably belong to the 
thick disk. Excluding these stars, a slight decreasing
trend of \feh\ with increasing $\Rm$ and age is found, but
a large scatter in \feh\ (up to 0.5 dex) is present at a
given age and $R_m$. Abundance ratios with respect to Fe
show, on the other hand, no significant scatter at a given \feh . The derived
trends of O, Mg, Si, Ca, Ti, Ni and Ba as a
function of \feh\ agree rather well with those of Edvardsson et al.,
but the overabundance of Na and Al for metal-poor stars found in their work is not confirmed.
Furthermore, the Galactic evolution
of elements not included in Edvardsson et al., K, V  and Cr, is studied.
It is concluded that the terms ``$\alpha$ elements" and ``iron-peak elements"
cannot be used to indicate production and evolution by specific
nucleosynthesis processes; each element seems to have a unique 
enrichment history.

 
\end{abstract}

\section{Introduction}

The chemical abundances of long-lived F and G main sequence stars,
combined with kinematical data and ages, provide a powerful way
to probe the chemical and dynamical evolution of the Galaxy.

 As far as the disk stars are concerned, many general trends have
been discovered during the past decades. Most notable results are
correlations of metallicity with age, Galactocentric distance, 
and vertical distance from the Galactic plane based on
photometric or low-resolution observations (e.g. Eggen et
al. \cite{Eggen62}; Twarog \cite{Twarog80}). 
In addition, the abundance patterns for some elements have been 
derived for small samples of stars: oxygen and $\alpha$ elements
relative to iron vary
systematically from overabundances at $\feh \simeq -1.0$ to a solar ratio
at $\feh \simeq 0.0$, while most iron-peak elements follow iron
for the whole metallicity range of the disk. These results have
provided important constraints on chemical evolution models
for the Galactic disk. 

With improved observation and analysis
techniques, which make it possible to study the Galactic chemical
evolution (GCE)
in detail, some old conclusions have, however, been challenged and
new questions have arisen. Particularly important is
the detailed abundance analysis of 189 F and G dwarfs with
$-1.1<\feh<0.25$ by Edvardsson et al. (\cite{Edvardsson93a}, hereafter EAGLNT). 
The main results from this work may be summarized
as follows: (1) There are no tight relations
between age, metallicity and kinematics of disk
stars, but substantial dispersions imposed on weak statistical
trends. (2)
There exists a real scatter in the run of [$\alpha$/Fe] vs. [Fe/H]
possibly due to the mixture of stars with different origins. The
scatter seems to increase with decreasing
metallicity starting at $\feh \simeq -0.4$. Together with a possible
increase in the dispersion of $W_{\mathrm{LSR}}$ (the stellar velocity
perpendicular to the Galactic plane with respect to the Local Standard
of Rest, LSR) at this point, the result
suggests a dual model for disk formation. It is, however,
unclear if the transition at $\feh \simeq -0.4$ represents the
division between the thin disk and the thick disk. (3) A group of
metal-poor disk stars with $\Rm < 7$~kpc 
is found to have larger
[$\alpha$/Fe] values than stars with $\Rm > 9$ Kpc, indicating a
higher star formation rate (SFR) in the inner disk than that in the outer disk. Since
essentially all the oldest stars in EAGLNT have small $\Rm$, it is, however,
difficult to know upon which, $\Rm$ or age, the main dependence of
[$\alpha$/Fe] is. (4) At a given age and $\Rm$, the scatter in
[$\alpha$/Fe] is negligible while [Fe/H] does show a
significant scatter. The former implies that the products of
supernovae of different types are thoroughly mixed into the interstellar
medium (ISM) before significant star
formation occurs. Based on this, the large scatter in \feh\
may be explained by infall of
unprocessed gas with a characteristic mixing time much longer
than that of the gas from supernovae of different types. 
(5) The Galactic scatter may be different for individual $\alpha$
elements; [Mg/Fe] and [Ti/Fe] show a larger scatter at a given
metallicity than [Si/Fe] and [Ca/Fe]. It suggests
that individual $\alpha$ elements may have different origins. (6) A new
stellar group, rich in Na, Mg, Al, was found among the
metal-rich disk stars, suggesting additional synthesis sources
for these elements.

Given that the study of EAGLNT was based on a limited
sample of stars with
certain selection effects and that the analysis technique
induced uncertainties in the final abundances, some subtle
results need further investigation before they can provide
reliable constraints on theory. For example, it is somewhat
unclear if the different [$\alpha$/Fe] at a given metallicity
between the inner disk and the outer disk stars is real and if old
disk stars are always located in the inner disk. Moreover, recent work by
Tomkin et al. (\cite{Tomkin97}) argued against the existence of
NaMgAl stars. In addition, a number of
elements, which are highly interesting from a nucleosynthetic point
of view, were not included in the work of EAGLNT.

The present work, based on a
large differently selected sample of disk stars, aims at
exploring and extending the results of EAGLNT with improved analysis techniques.
Firstly, we now have more reliable atmospheric parameters. The
effective temperature is derived from the Str\"omgren $b-y$ color index using
a recent infrared-flux calibration and the surface gravity is based
on the Hipparcos parallax. About one hundred iron lines (instead of $\sim 30$
in EAGLNT) are used to
provide better determinations of metallicity and microturbulence.
Secondly, the abundance calculation is anchored at 
the most reliable theoretical or experimental oscillator strengths presently
available in the literature.
Thirdly, greater numbers of \Fetwo, \Sione\ and \Caone\ lines in
our study should allow better abundance
determinations, and new elements  (K, Sc, V, Cr and Mn) will give
additional information on Galactic evolution. Lastly, the stellar
age determination is based on new evolutionary tracks, and the
space velocity is derived from more reliable
distance and proper motion values.

In the following sections \ref{sec:obs} to \ref{sec:age}, we
describe the observations and methods of analysis in 
details and present the derived abundances, ages and kinematics.
The results are discussed in Sect.~\ref{sec:result} and compared
to those of EAGLNT. Two  elements, Sc and Mn, not included in
EAGLNT and represented by lines showing significant hyperfine
structure (HFS) effects, are discussed in a separate paper
(Nissen et al. \cite{Nissen99}), which
includes results for halo stars from Nissen \& Schuster (\cite{Nissen97}).                               

\section{Observations}
\label{sec:obs}
\subsection{Selection of stars} 
\label{sec:selection}

The stars were selected from the $uvby-\beta$ photometric catalogues
of Olsen (\cite{Olsen83}, \cite{Olsen93}) according to the
criteria of $5800 \leq \teff \leq
6400~\mathrm{K}$, $4.0 \leq \logg \leq 4.5$ and $ -1.0 \leq \feh \leq +0.3 $
with approximately equal numbers of stars in every metallicity
interval of 0.1 dex. In this selection, 
the temperature was determined from the $b-y$ index with the calibration of
Magain (\cite{Magain87}), gravity was calculated from the
$c_1 $ index as described in
EAGLNT, and metallicity was derived from the $ m_1 $ index using
the calibrations of Schuster \& Nissen (\cite{Schuster89}). The
later redeterminations of
the temperature with the calibration of Alonso et al. (\cite{Alonso96}) and the gravity
from Hipparcos parallax lead to slight deviations from the selection
criteria for some stars.

Based on the above selection, 104 F and G stars were
observed, but 3 high-rotation ($V \sin i \geq 25 \kmprs $) stars and 9 double-line
spectroscopic binaries were excluded from the sample.  
Another 11 stars have radial velocity dispersions higher than the
measurement error of the CORAVEL survey. 
\object{HD\,106516A} and \object{HD\,97916} are suspected binaries (Carney
et al. \cite{Carney94}), and \object{HD\,25998} and \object{HD\,206301} are possibly
variables (e.g. Petit \cite{Petit90}; Morris \& Mutel \cite{Morris88}).
These 15 stars (marked in the column ``Rem'' of Table 3 
are being
carefully used in our study. The remaining stars are considered as
single stars, but are checked for differences in
iron abundances between \Feone\ and \Fetwo\ lines using gravities
from Hipparcos parallaxes  
as suggested by Fuhrmann (\cite{Fuhrmann98}). As described later,
additional 2 stars were excluded during the analysis and thus the
sample contains 90 stars for the final discussion and conclusions.

\subsection{Observations and data reduction}

The observations were carried out with the Coud\'{e} Echelle
Spectrograph attached to the 2.16m telescope at Beijing
Astronomical Observatory (Xinglong, P.R. China). The detector was a
Tek CCD ($ 1024\times 1024$ pixels with $24\times 24~\mu m^{2}$ each in
size). The red
arm of the spectrograph with a 31.6 grooves/mm grating
was used in combination with a prism as cross-disperser, providing a
good separation between the echelle orders.  With a 0.5 mm slit
(1.1 arcsec), the resolving power was of the order of 40\,000 in the
middle focus camera system.

\begin{figure}
\resizebox{\hsize}{!}{\includegraphics{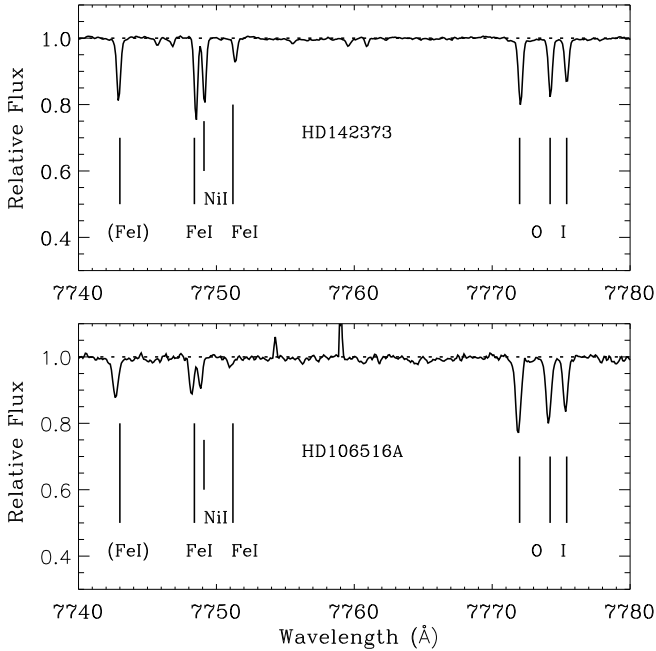}}
\caption{Examples of spectra obtained with the 2.16m telescope at
Xinglong Station for \object{HD\,142373} ($\teff=5920 $~K,
$\logg=4.27$, $\feh=-0.39 $) with a high S/N $\sim$ 400 and
\object{HD\,106516A} ($\teff=6135 $~K, $\logg=4.34$,
$\feh=-0.71$) with a relatively low S/N $\sim$ 160.}
\label{fig:sp142373}
\end{figure}

The program stars were observed during three runs:
March 21-27, 1997 (56 stars), October 21-23, 1997 (27 stars) and
August 5-13, 1998 (21 stars). 
The exposure time was chosen in order to
obtain a signal-to-noise ratio of at least 150 over the entire spectral
range. Most bright stars have S/N $\sim$ 200 -- 400. Figure~\ref{fig:sp142373} 
shows the spectra in the region of the oxygen triplet for two representative
stars \object{HD\,142373} and \object{HD\,106516A}. 
In addition, the solar flux spectrum as reflected from the
Moon was observed with a S/N $\sim$ 250 and used as one of the ``standard'' stars in
determining oscillator strengths for some lines (see Sect.~\ref{subsec:gf}).

The spectra were reduced with standard MIDAS routines for order identification, background
subtraction, flat-field correction, order extraction and
wavelength calibration. Bias, dark current and scattered light
corrections are included in the background subtraction. If an
early B-type star could be observed close to the program stars,
it was used instead of the flat-field in order to remove interference fringes more
efficiently. 
The spectrum was then normalized by a continuum function determined
by fitting a spline curve to a set of pre-selected continuum
windows estimated from the solar atlas. Finally,
correction for radial velocity shift, measured from at least 20
lines, was applied before the measurement of equivalent widths.
\subsection{Equivalent widths and comparison with
EAGLNT}
\label{subsec:ewcom}
The equivalent widths were measured by three methods: direct
integration, Gaussian and Voigt function fitting, depending on
which method gave the best fit of the line profile. Usually, weak lines
are well fitted by a Gaussian, whereas stronger lines in which
the damping wings contribute significantly to their equivalent widths
need the Voigt function to reproduce their profiles. If
unblended lines are well separated from nearby lines, direct integration
is the best
method. In the case of some intermediate-strong lines, weighted averages
of Gaussian and Voigt fitting were adopted.

The accuracy of the equivalent widths is estimated by comparing them
to the independent measurements by EAGLNT for 25 stars
in common. Five of them were observed at the ESO
Observatory ($R \sim 60\,000$, S/N $\sim$ 200) and 23 were observed
at the McDonald Observatory ($R \sim 30\,000$, S/N $\sim$
200 -- 500).
\begin{figure}
\resizebox{\hsize}{!}{\includegraphics{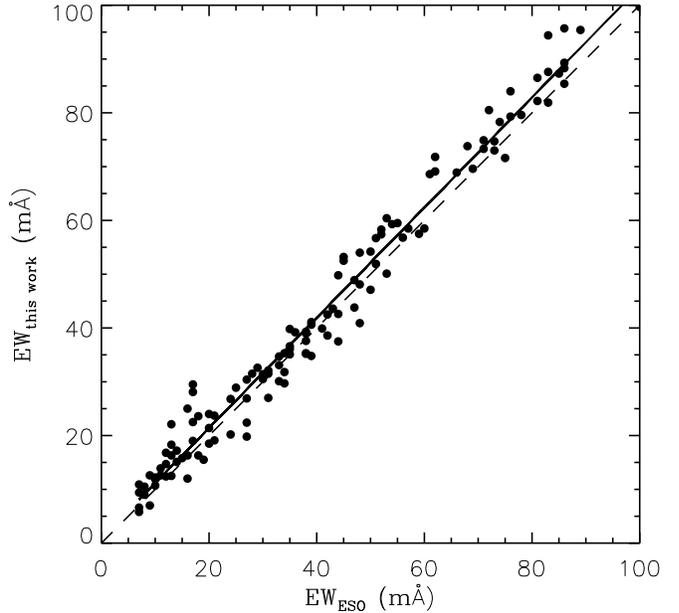}}
\caption{A comparison of equivalent widths   
measured in this work with ESO data in EAGLNT for 5 stars in common.
The thick line is the linear fit to the points, whereas the dashed
line is the one-to-one relation.}
\label{fig:esoew}
\end{figure}
The systematic difference between the two sets of measurements is
small and a linear least squares fitting gives:
\begin{eqnarray}
EW_{\mathrm{Xl}}&=&1.025 \left(\pm0.012\right)  EW_{\mathrm{ESO}} + 0.89 \left(\pm0.56\right) \,\,   
(m{\mbox{\AA}}) \nonumber\\
EW_{\mathrm{Xl}}&=&1.083 \left(\pm0.006\right) EW_{\mathrm{McD}} - 0.94 \left(\pm0.28\right)  \,\, 
(m{\mbox{\AA}}) \nonumber
\end{eqnarray}
The standard deviations around the two relations are
3.8 m\AA\ (for 129 lines in common with ESO) and 4.3 m\AA\ (for
575 lines in common with McDonald). Given that the error of
the equivalent widths in EAGLNT is around 2 m\AA, we estimate an
rms error of about 3 m\AA\ in our equivalent widths. As shown
by the comparison of  our equivalent widths with ESO data in
Fig.~\ref{fig:esoew}, the equivalent widths below 50 m\AA\ are
consistent with the one-to-one relation. The deviations for
the stronger lines may be due to the fact that all lines in EAGLNT were
measured by Gaussian fitting, which leads to an underestimate of
equivalent widths for intermediate-strong lines because of neglecting
their wings. We conclude that the Xinglong data may be more reliable than
the EAGLNT data for lines in the range of $50 < EW < 100$~m\AA .

\section{Stellar atmospheric parameters}
\label{sec:para}
\subsection{Effective temperature and metallicity}
The effective temperature was determined from the Str\"omgren indices
($b-y$ and $c_1$) and \feh\ using the calibration of
Alonso et al. (\cite{Alonso96}).
If the color excess $E(b-y)$, as calculated from the $H_{\beta}$ index
calibration by Olsen (\cite{Olsen88}), is larger than 0.01, then
a reddening correction was applied.

The metallicity, required in the input for temperature and
abundance calculation, was first derived from the Str\"omgren $m_1$ index
using the calibrations
of Schuster \& Nissen (\cite{Schuster89}). But the spectroscopic
metallicity obtained later was used to iterate the whole
procedure.

The errors of the photometric data are $\sigma(b-y) = 0.004$ and $\sigma(c_1)
= 0.008$ according to Olsen (\cite{Olsen93}). Adopting
$\sigma(\feh)$ = 0.1 from the spectroscopic analysis,
the statistical
error of \teff\ is estimated to about $\pm 50$~K. Considering a possible
error of $\pm$50~K in the calibration, the error in temperature could reach
$\pm 70$~K. We do not adopt the excitation
temperature, determined from a consistent abundance derived from
\Feone\ lines with different excitation potentials, because
errors induced by incorrect damping parameters (Ryan \cite{Ryan98})
or non-LTE effects can be strongly
dependent on excitation potential, leading to an error in effective
temperature as high as 100~K.

\subsection{Gravity}
\label{subsec:logg}
In most works, gravities are determined from the abundance
analysis by requiring that \Feone\ and \Fetwo\ lines give the
same iron abundance. But it is
well known that the derivation of iron abundance from \Feone\
and \Fetwo\ lines may be affected by many factors such as
unreliable oscillator strengths, possible non-LTE
effects and uncertainties in the temperature structure of the model
atmospheres. 
From the Hipparcos parallaxes, we can determine more
reliable gravities using the relations:
\begin{eqnarray}
 \log \frac{g}{g_{\odot}}&=&\log \frac{\mathcal M}{{\mathcal M}_{\odot}} 
+4 \log \frac{\teff}{\teff_{\odot}}+
0.4 \left(M_{\mathrm{bol}}-M_{\mathrm{bol},\odot}\right)  \label{eq:logg} 
\end{eqnarray}
and
\begin{eqnarray}
 M_{\mathrm{bol}}&=& V+BC+5 \log \pi + 5,\label{eq:Mv}
\end{eqnarray}
where, ${\mathcal M}$ is the stellar mass, $M_{\mathrm{bol}}$ the
absolute bolometric magnitude, $V$ the visual magnitude, $BC$
the bolometric correction, and $\pi$ the parallax. 

The parallax is taken from the Hipparcos Satellite observations
(ESA \cite{ESA97}). For most program stars, the relative error in
the parallax is of the order of 5\%. Only two stars in our sample
have errors larger than 10\%. From these accurate parallaxes,
stellar distances and absolute magnitudes were obtained. Note, however, that our sample includes some binaries, for which the
absolute magnitude from the Hipparcos parallax could be significantly in error.
An offset of $-0.75$~mag will be introduced for a
binary with equal components through the visual magnitude in
Eq.~(\ref{eq:Mv}).
Thus, we also calculated absolute magnitudes from the photometric
indices $\beta$ and $c_{1}$ using the relations found by
EAGLNT. Although the absolute magnitude of a binary derived by the
photometric method is also not very accurate due to different
spectral types and thus different flux distributions of the
components, it may be better than the value from the parallax
method. Hence, for a few stars with large differences in absolute
magnitudes between the photometry and parallax determination, we
adopt the photometric values.

The bolometric correction was interpolated from the new BC grids of
Alonso et al. (\cite{Alonso95}) determined from line-blanketed flux
distributions of ATLAS9 models. It is noted that the zero-point of
the bolometric correction adopted by Alonso et al.,
$BC_{\mathrm{\odot}}=-0.12$, is not consistent with the bolometric magnitude
of the Sun, $M_{\mathrm{bol},\odot}$=4.75, recently recommended by the
IAU (\cite{IAU99}). But the gravity determination from the Eq.~(\ref{eq:logg}) only
depends on the $M_{\mathrm{bol}}$ difference between the stars and the Sun
and thus the zero-point is irrelevant.

The derivation of mass is described in Sect.~\ref{sec:age}.
The estimated error of 0.05 $M_{\odot}$ in mass corresponds to an
error of 0.03 dex in gravity,
while errors of $0.05$~mag in BC and 70~K in temperature each leads to
an uncertainty of 0.02 dex in \logg. The largest uncertainty of
the gravity comes from the parallax. A typical relative error of 5\%
corresponds to an error of 0.04 dex in \logg. In total, the error
of \logg\ is less than 0.10 dex.

The surface gravity was also estimated from the Balmer
discontinuity index $c_{1}$ as described in EAGLNT. We find a
small systematical shift (about 0.1 dex) between the two sets of
\logg, with lower gravities
from the parallaxes. There is no corresponding shift between
$M_{V}$(par) and $M_{V}$(phot). The mean deviation is 0.03 mag only,
which indicates that the systematic deviation in \logg\
comes from the gravity calibration in EAGLNT.

\subsection{Microturbulence}
\label{subsec:Vt}
The microturbulence, $\xi_{t}$, was determined from the abundance
analysis by requiring a zero slope of [Fe/H] vs. EW.
The large number of \Feone\ lines in this study enables us
to choose a set of lines with accurate oscillator strengths,
similar excitation potentials ($\chi_{low} \geq 4.0$ eV) and a
large range of equivalent widths (10 -- 100 m\AA ) for the
determination. With this selection, we hope to reduce the errors
from oscillator strengths and potential non-LTE effects for \Feone\ lines
with low excitation potentials. The error of the microturbulence
is about 0.3 \kmprs .

The relation of $\xi_{t}$ as a function of \teff\ and \logg\ 
derived by EAGLNT corresponds to about 0.3 \kmprs\
lower values than those derived from our spectroscopic analysis. 
No obvious dependence of the difference on temperature,
gravity and metallicity can be found. In particular, the value for the Sun in
our work is 1.44 \kmprs, also 0.3 \kmprs\ higher than
the value of 1.15 found from the EAGLNT relation. The difference in
$\xi_{t}$ between EAGLNT and the present work is probably related to
the difference in equivalent widths of intermediate-strong lines discussed in
Sect.~\ref{subsec:ewcom}. EAGLNT measured these lines by fitting
a Gaussian function and hence underestimated their equivalent
widths, leading to a lower microturbulence.

Finally, given that the atmospheric parameters were not determined
independently, the whole procedure of deriving \teff,  \logg,
\feh\ and $\xi_{t}$ was iterated to consistency. The atmospheric
parameters of 90 stars are presented in Table 3. 
The uncertainties of the
parameters are: $\sigma(\teff) = 70$~K, $\sigma(\logg) = 0.1$,
$\sigma(\feh) = 0.1$, and $\sigma(\xi_{t})=0.3$ \kmprs .

\section{Atomic line data} \label{sec:ldata}
\subsection{Spectral lines}

All unblended lines with symmetric profiles having equivalent
widths larger than 20 m\AA\ in the solar atlas (Moore et al. \cite{Moore66}) 
were cautiously selected. The equivalent width limit ensures that
lines are not disappearing in the most metal-poor disk stars at
$\feh = -1.0$. Given that very weak lines would lead to an
increase of random errors in the abundance determination and
that too strong lines are very sensitive to
damping constants, only weak and intermediate-strong lines with
$3 < EW < 100$~m\AA\ in the stellar spectra were
adopted in our abundance analysis except for potassium, for which only
one line ($\Kone$ $\lambda$7699) with an equivalent width range of 50 -- 190~m\AA ,
is available.

\subsection{Oscillator strengths} 
\label{subsec:gf}

Due to the large number of measurable lines in the spectra,
\Feone\ lines  were used for microturbulence determination and
temperature consistency check. Hence, careful selection of
oscillator strengths for them is of particular importance. Of
many experimental or theoretical calculations of oscillator
strengths for \Feone\ lines, only three sources with precise $gf$
values were chosen. They are: Blackwell et al. (\cite{Blackwell82b}; \cite{Blackwell82c}),
O'Brian et al. (\cite{OBrian91}) and Bard \& Kock (\cite{Bard91})
or Bard et al. (\cite{Bard94}). The
agreements between these sources are very satisfactory, and thus 
mean log$gf$ values were adopted if oscillator strengths are
available in more than one of the three sources. A few oscillator
strengths with large differences between these sources were excluded.

References for other elements are: \Oone\ (Lambert
\cite{Lambert78}), \Naone\ (Lambert \& Warner \cite{Lambert68}),
\Mgone\ (Chang \cite{Chang90}), \Alone\ (Lambert \& Warner \cite{Lambert68}), \Sione\
(Garz \cite{Garz73}), \Kone\ ( Lambert \& Warner \cite{Lambert68}), \Caone\ (Smith \& Raggett
\cite{Smith81}), \Tione\ (Blackwell et al. \cite{Blackwell82a}; \cite{Blackwell86a}), \Vone\
(Whaling et al. \cite{Whaling85}), \Crone\ (Blackwell et
al. \cite{Blackwell86b}), \Fetwo\ (Bi\'{e}mont et
al. \cite{Biemond91}; Hannaford et al. \cite{Hannaford92}), \Nione\
(Kostyk \cite{Kostyk82}; 
Wickliffe \& Lawler
\cite{Wickliffe97}) and \Batwo\ (Wiese \& Martin \cite{Wiese80}).

These experimental or theoretical $gf$ values were inspected to
see if they give reliable abundances by evaluating the deviation
between the abundance derived from a given line and the mean
abundance from all lines of the same species. A significant mean
deviation in the same direction for all stars (excluding the
suspected binaries) were used to correct the $ gf$ value. Lines
with large deviations in different directions for different stars
were discarded from the experimental or theoretical log$gf$ list.
The two sets of values are presented in columns ``abs.'' and
``cor.'' in Table 4. 
Based on these $gf$ values,
we derived abundances for 10 ``standard'' stars:
\object{HD\,60319}, \object{HD\,49732},
\object{HD\,215257}, \object{HD\,58551}, \object{HD\,101676},
\object{HD\,76349}, \object{HD\,58855}, \object{HD\,22484}, the Sun and
\object{HD\,34411A} (in metallicity order).
Oscillator strengths for lines with unknown $gf$ values were then
determined from an
inverse abundance analysis of the above 10 stars, which are
distributed in the metallicity range $ -0.9 \leq
\feh \leq 0.0 $ dex and were observed at high S/N $\sim$ 250 -- 400.
Generally, the $gf$ values for a given line from different ``standard''
stars agree well, and a mean value (given in the column ``dif'' in 
Table 4 
was thus adopted.

\subsection{Empirical enhancement factor}

It has been recognized for a long time that line broading derived
from Uns\"{o}ld's (\cite{Unsold55}) approximation to Van der Waals
interaction is too
weak, and an enhancement factor, $E_{\gamma}$, should be applied to the damping
parameter, $\gamma$. Usually, enhancement factors of \Feone\ lines with
excitation potential of the lower energy level ($\chi_{low}$) less
than 2.6 eV are taken from the empirical calibration by Simmons \& Blackwell (\cite{Simmons82}). For
\Feone\ lines with $\chi_{low}>2.6$ eV, $E_{\gamma}=1.4$ is
generally used in abundance analysis. Recently,
Anstee \& O'Mara's (\cite{Anstee95}) computed the broadening cross
sections for s-p and p-s transitions
and found that $E_{\gamma}$ should be $2.5\sim3.2$ for lines
with $\chi_{low}>3.0$ eV, whereas lines with $\chi_{low}<2.6$ eV
have broadening cross sections more consistent with Simmons \&
Blackwell's (\cite{Simmons82}) work.

Following EAGLNT and other works, we adopted Simmons \&
Blackwell's (\cite{Simmons82}) $E_{\gamma}$ for \Feone\ lines
with  $\chi_{low}<2.6$ eV and
$E_{\gamma}=1.4$ for the remaining \Feone\ lines, while
a value of 2.5 was applied for \Fetwo\ lines as suggested by
Holweger et al. (\cite{Holweger90}).
Enhancement factors for \Naone, \Sione, \Caone\
and \Batwo\ were taken from EAGLNT (see references therein).
Finally, a value of 1.5 was adopted for the \Kone, \Tione\ and \Vone\ lines considering their 
low excitation potentials, and a factor of 2.5 was applied to
the remaining elements following M\"ackle et al. (\cite{Mackle75}).
The effects of changing these values by 50\% on the derived
abundances are discussed in Sect.~\ref{subsec:abuerra}.

The atomic line data are given in Table 4. 

\section{Abundances and their uncertainties}
\label{sec:abu}

\subsection{Model atmospheres and abundance calculations}
The abundance analysis is based on a grid of flux constant,
homogeneous, LTE model atmospheres, kindly supplied by
Bengt Edvardsson (Uppsala).  The models were computed with
the MARCS code using the updated continuous opacities by Asplund et
al. (\cite{Asplund97}) including UV line blanketing by millions
of absorption lines and many molecular lines. 

The abundance was calculated with the program EQWIDTH
(also made available from the stellar atmospheric group in Uppsala)
by requiring that the calculated
equivalent width from the model should match the observed value.
The calculation includes natural broadening, thermal
broadening, van der Waals damping, and the microturbulent Doppler
broadening. The mean abundance was derived from all available lines by
giving equal weight to each line. Finally, solar abundances,
calculated from the Moon spectrum, were used to derive stellar abundances
relative to solar values (Table 5). 
Such differential abundances are generally more reliably than 
absolute abundances because many systematic errors nearly cancel out. 

\subsection{Uncertainties of abundances}
\label{subsec:abuerra}
There are two kinds of uncertainties in the abundance determination: 
one acts on individual lines, and includes random errors of equivalent widths, oscillator strengths,
and damping constants; another acts on the whole set of lines
with the main uncertainties coming from the atmospheric parameters.

\subsubsection{Errors from equivalent widths and atomic data}
The comparison of equivalent widths in Sect.~\ref{subsec:ewcom}
indicates that the typical uncertainty of the equivalent width 
is about 3 m\AA, which leads to an error of about 0.06 dex in the
elemental ratio X/H derived from a single line with an equivalent
width around 50 m\AA.
For an element represented by N lines, the error is decreased by a factor
$\sqrt{N}$. In this way, the errors from equivalent widths
were estimated for elements with only one or a few lines.
Alternatively, the scatter of the deduced abundances from a 
large number of lines with reliable oscillator strengths gives
another estimate of 
the uncertainty from equivalent widths. With over 100 \Feone\ 
lines for most stars, the scatter varies somewhat from star to
star with a mean value of 0.07 dex, corresponding to an error
of 0.007 dex in \feh .
Other elements with significant numbers of lines, such as Ca, Ni and Si, have
even smaller mean line-to-line scatters.

The uncertainties in atomic data are more difficult to
evaluate. But any
error in the differential abundance caused by errors in the
$gf$ values is nearly excluded due to the correction of some experimental
or theoretical $gf$ values and the adoption of
mean $gf$ values from 10 ``standard'' stars.
Concerning the uncertainties in the damping constants, we have
estimated their effects by increasing the adopted enhancement
factors by 50\%. The microturbulence was
accordingly adjusted because of the coupling between the two parameters.
The net effect on the differential abundances with respect to the
Sun is rather small as seen from Table~\ref{tb:abuerr}.

\subsubsection{Consistency check of atmospheric parameters}
As a check of the photometric temperature, the derived iron abundance
from individual \Feone\ lines was studied as 
a function of the excitation potential. To
reduce the influence of microturbulence, only lines with
equivalent widths less than 70 m\AA\ were included.
A linear least squares fit to the
abundance derived from each line vs. low excitation potential
determines the slope in the relation
$\feh = a + b \cdot \chi_{low}$. The mean slope  coefficient for
all stars is $b=0.004 \pm 0.013 $. 
There is only a very small (if any) dependence of
$b$ on effective temperature, surface gravity or metallicity. A
suspected binary, \object{HD\,15814},
has a very deviating slope coefficient ($b=-0.056$) and is
excluded from further analysis.

\begin{figure}
\resizebox{\hsize}{!}{\includegraphics{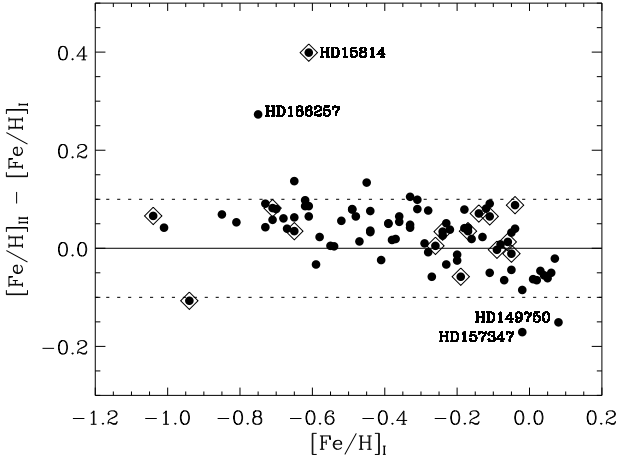}}
\caption{Difference in iron abundances derived from \Feone\ and
  \Fetwo\ lines vs. [Fe/H] with suspected
binaries marked by a square around the
filled circles.}
\label{fig:fe2fe1}
\end{figure}

The agreement of iron abundances derived from \Feone\ and \Fetwo\
lines is satisfactory when gravities
based on Hipparcos parallaxes are used (see Fig.~\ref{fig:fe2fe1}).
The deviation is less than 0.1 dex for most stars with a mean value
of $-0.009 \pm 0.07$ dex.
From $\feh=0.0$ to $\feh=-0.5$, the mean deviation ($\feh_{\mathrm{II}}-\feh_\mathrm{I}$) 
seems, however,
to increase by about 0.1 dex in rough agreement with predictions
from non-LTE computations (see Sect.~\ref{subsec:NLTE}).

The deviation in iron abundances based on \Feone\ and \Fetwo\ abundance 
provides a way to
identify binaries and to estimate the influence of the component on the 
primary. The suspected binaries are marked with an additional square
around the filled circles in Fig.~\ref{fig:fe2fe1}. It shows that
there is no significant influence from the
component for these binaries except in the case of
\object{HD\,15814}, which was already excluded on the basis of it's
$b$-coefficient in the excitation equilibrium of \Feone\ lines. 
Thus, the  other possible binaries were included in our analysis.
It is, however, surprising that \object{HD\,186257} show a higher iron
abundance based on \Fetwo\ lines than that from \Feone\ lines with
a deviation as
large as 0.28 dex. We discard this star in the final analysis and
thus have 90 stars left in our sample.

\subsubsection{Errors in resulting abundances}
\label{subsec:abuerr} 

Table~\ref{tb:abuerr} shows the effects on the
derived abundances of a change by
70~K in effective temperature, 0.1 dex in gravity, 0.1 dex in metallicity,
and 0.3 $\kmprs$ in microturbulence, along with errors from equivalent
widths and enhancement factors, for two representative
stars. 

\begin{table}
\setlength{\tabcolsep}{0.1cm}
\caption[]{Abundance errors. The last column gives the total error assuming that the 
individual errors are uncorrelated}
\label{tb:abuerr}
\begin{tabular}{lrrrrrrr}
\noalign{\smallskip}
\hline
\noalign{\smallskip}
\multicolumn{8}{l}{HD142373~$\teff=5920~\logg=4.27~\feh = $--$0.39~\xi_t=1.48$} \\
\hline
\noalign{\smallskip}
 &$\frac{\sigma_{EW}}{\sqrt{N}}$&$\Delta \teff$ &$\Delta \logg $& $\Delta \ffe $  & $\Delta \xi_t$ & $\Delta E_{\gamma}$& $\sigma_{tot}$\\
       &     & +70K  & +0.1  & +0.1  & +0.3  & 50\%  \\
\noalign{\smallskip}
\hline
$\Delta \ffe$      & .009&  .048& $-$.005&  .001& $-$.040&  .015 & .065\\[1mm]
$\Delta \ffe_{II}$ & .030& $-$.012&  .031&  .018& $-$.046& .007 & .067\\[1mm]
$\Delta \ofe$      & .035& $-$.104&  .024&  .004&  .020& .013& .115\\[1mm]
$\Delta \nafe$     & .042& $-$.017&  .004& $-$.001&  .034& .011& .058\\[1mm]
$\Delta \mgfe$     & .042& $-$.018& $-$.005&  .005&  .019& .015 & .052\\[1mm]
$\Delta \alfe$     & .035& $-$.023&  .003& $-$.001&  .033&  .015& .055\\[1mm]
$\Delta \sife$     & .015& $-$.026&  .005&  .003&  .025&  .015& .042\\[1mm]
$\Delta \cafe$     & .023& $-$.002& $-$.008&  .002& $-$.005& .011& .028\\[1mm]
$\Delta \tife$  &    .024&  .017&  .004&  .000&  .023& .009  & .039\\[1mm]
$\Delta \vfe$      & .035&  .019&  .004&  .000&  .031& .011 & .051\\[1mm]
$\Delta \crfe$     & .035& $-$.007&  .001&  .000&  .025&  .015& .053\\[1mm]
$\Delta \nife$     & .011&  .002&  .004&  .003&  .013&  .023& .029\\[1mm]
$\Delta \bafe$     & .060&  .039& $-$.008&  .005& $-$.041&  .012& .084\\[1mm]
$\Delta \kfe$      & .060&  .013& $-$.024&  .013& $-$.029&  .012& .074\\[1mm]
\noalign{\smallskip}
\hline
\end{tabular}

\setlength{\tabcolsep}{0.1cm}
\begin{tabular}{lrrrrrrr}
\noalign{\smallskip}
\hline
\noalign{\smallskip}
\multicolumn{8}{l}{ HD106516~$\teff=6135~\logg=4.34~\feh = $--$0.71~\xi_t=1.48$} \\
\hline
\noalign{\smallskip}
 &$\frac{\sigma_{EW}}{\sqrt{N}}$&$\Delta \teff$ &$\Delta \logg $& $\Delta \ffe $  & $\Delta \xi_t$ & $\Delta E_{\gamma}$& $\sigma_{tot}$\\
& &   +70 K  & +0.1  & +0.1  & +0.3  & 50\% &   \\
\noalign{\smallskip}
\hline
$\Delta \ffe_{I}$& .012&  .042& $-$.003&  .006& $-$.029& .022& .057\\[1mm]
$\Delta \ffe_{II}$& .023&  .000&  .032&  .008& $-$.025&.016& .050\\[1mm]
$\Delta \ofe$ & .042& $-$.079&  .017& $-$.003&  .007& $-$.003& .091\\[1mm]
$\Delta \nafe$& .042& $-$.018&  .002& $-$.003&  .026&  .013& .054\\[1mm]
$\Delta \mgfe$& .042& $-$.014& $-$.009&  .001&  .012& .022& .052\\[1mm]
$\Delta \sife$& .019& $-$.021&  .004& $-$.001&  .019&.023  & .041\\[1mm]
$\Delta \cafe$& .015& $-$.001& $-$.009&  .002& $-$.009& .014& .024\\[1mm]
$\Delta \tife$& .030&  .010&  .003&  .000&  .021& .010& .039\\[1mm]
$\Delta \vfe$ & .042&  .014&  .003&  .000&  .025& .006& .051\\[1mm]
$\Delta \crfe$& .035& $-$.006&  .001& $-$.001&  .018&  .034& .052\\[1mm]
$\Delta \nife$& .012& $-$.001&  .002& $-$.001&  .013&  .029& .034\\[1mm]
$\Delta \bafe$& .042&  .041& $-$.012&  .008& $-$.047& .008& .077\\[1mm]
$\Delta \kfe$ & .060&  .016& $-$.026&  .010& $-$.039& .024&.082\\[1mm]
\noalign{\smallskip}
\hline
\end{tabular}
\end{table}

It is seen that the relative
abundances with respect to iron are quite insensitive to variations
of the atmospheric parameters.  One exception is [O/Fe] due to
the well known fact that the oxygen abundance derived from the
infrared triplet has an opposite
dependence on temperature to that of the iron abundance.
After rescaling of our
oxygen abundances to results from the forbidden line at $\lambda 6300$
(see next section) the error is somewhat reduced.
Therefore, the error for [O/Fe] in Table~\ref{tb:abuerr} might be overestimated.

In all, the uncertainties of the atmospheric parameters give errors
of less than 0.06 dex in the resulting [Fe/H] values and 
less than 0.04 dex in the relative abundance ratios. For an elemental abundance
derived from many lines, this is the dominant error,
while for an abundance derived from a few lines, the uncertainty in the
equivalent widths may be more significant.
Note that the uncertainties of equivalent widths
for V and Cr (possibly also Ti) might be underestimated given that their lines
are generally weak in this work. In addition,
with only one strong line for the K abundance determination, the
errors from equivalent widths, microturbulence and atomic line data
are relatively large.

Lastly, we have explored the HFS effect on one \Alone\ line at 
$\lambda$6698 , one \Mgone\ line at $\lambda$5711, and two \Batwo\ lines at
$\lambda$6141 and $\lambda$6496. The HFS data are taken from three sources: Biehl (\cite{Biehl76})
for Al, Steffen (\cite{Steffen85}) for Mg and  Fran\c{c}ois (\cite{Francois96}) for
Ba. The results indicate that the HFS effects are very small for
all these lines with a value less than 0.01 dex.

\subsection{Non-LTE effects and inhomogeneous models} \label{subsec:NLTE}

The assumption of LTE and the use of homogeneous model atmospheres
may introduce systematic errors, especially on the slope of various
abundance ratios [X/Fe] vs. [Fe/H]. These problems were discussed at quite
some length by EAGLNT. Here we add some remarks based on recent
non-LTE studies and computations of 3D hydrodynamical model atmospheres.

Based on a number of studies, EAGLNT concluded that the maximum non-LTE
correction of [Fe/H], as derived from \Feone\ lines, is 0.05 to 0.1 dex
for metal-poor F and G disk dwarfs.  Recently, Th\'{e}venin \& Idiart
(\cite{Thevenin99})
computed non-LTE corrections on the order of 0.1 to 0.2 dex at $\feh = -1.0$.
Fig.~\ref{fig:fe2fe1} suggests that the maximium correction to $\feh$
derived from \Feone\ lines is around 0.1 dex, but we emphasize that this
empirical check may depend on the adopted $\teff$ calibration as a function
of $\feh$.

The oxygen infrared triplet lines are suspected to be affected by
non-LTE formation,
because they give systematically higher abundances than forbidden
lines.  
Recent work by Reetz
(\cite{Reetz98}) indicates that non-LTE effects are insignificant ($< 0.05$~dex)
for metal-poor and cool stars, but become important for warm and
metal-rich stars.
For stars with $\feh > -0.5$ and $\teff > 6000$~K in our sample,
non-LTE effects could reduce the oxygen abundances by 0.1-0.2
dex. For this reason, we use Eq. (11) of EAGLNT to scale
the oxygen abundances derived from the
infrared triplet to those derived by Nissen \& Edvardsson (\cite{Nissen92})
from the forbidden [\Oone] $\lambda$6300.
 
The two weak \Naone\ lines ($\lambda 6154$ and $\lambda 6160$)
used for our Na abundance determinations, are only marginally affected by
deviations from LTE formation (Baum\"{u}ller et al. \cite{Baumuller98}).
The situation for Al may, however, be different. The non-LTE analysis
by Baum\"{u}ller \& Gehren (\cite{Baumuller97})
of one of the \Alone\ lines used in the present work ($\lambda 6698$)
leads to about 0.15 dex higher Al abundances for the metal-poor disk dwarfs
than those calculated from LTE.  No non-LTE study
for the other two lines used in the present work is
available. We find, however, that the derived Al abundances depend
on $\teff$ with
lower [Al/Fe] for higher temperature stars. This may be due to
the neglect of non-LTE effects in our work. Hence, we suspect that 
the trend of [Al/Fe] vs. \feh\ could be seriously affected by non-LTE effects.

The recent non-LTE analysis of neutral magnesium in the solar atmosphere
by Zhao et al. (\cite{Zhao98}) and in metal-poor stars by Zhao \& Gehren
(\cite{Zhao99}) leads to non-LTE corrections of 0.05 dex for the Sun
and 0.10 dex for a $\feh = -1.0$ dwarf, when the abundance of Mg is derived
from the $\lambda 5711$ \Mgone\ line. Similar corrections are obtained 
for some of the other lines used in the present work. Hence, we conclude
that the derived trend of [Mg/Fe] vs. [Fe/H] is not significantly
affected by non-LTE.

The line-profile analysis of the \Kone\ resonance line at $\lambda$7699 by
Takeda et al. (\cite{Takeda96}) shows that the non-LTE correction is 
$-0.4$~dex for the Sun and $-0.7$~dex for Procyon. There are no
computations for metal-poor stars, but given the very large corrections for
the Sun and Procyon one may expect that the slope of [K/Fe] vs. [Fe/H] could be
seriously affected by differential non-LTE effects between the Sun and 
metal-poor stars.

The non-LTE study of Ba lines by Mashonkina et al. (\cite{Mashonkina99}),
which includes two of our three \Batwo\ lines ($\lambda$5853 and $\lambda$6496),
give rather small corrections ($<0.10$ dex) to the LTE abundances, and the 
corrections are very similar for solar metallicity and $\feh \simeq -1.0$
dwarfs. Hence, [Ba/Fe] is not affected significantly.

In addition to possible non-LTE effects, the derived abundances  may also
be affected by the representation of the stellar atmospheres by 
plane-parallel, homogeneous models. The recent 3D hydrodynamical model
atmospheres of metal-poor stars by Asplund et al. (\cite{Asplund99})
have substantial lower temperatures in the upper photosphere
than 1D models due to the dominance of adiabatic cooling over 
radiative heating. Consequently, the iron abundance
derived from \Feone\ lines in a star like 
\object{HD\,84937} ($\teff \simeq 6300$~K, $\logg \simeq 4.0$ and 
$\feh \simeq -2.3$) is 0.4~dex lower than  the value based on a 
1D model. Although the effect will be smaller in a $\feh \simeq -1.0$ star,
and the derived abundance ratios are not so sensitive to
the temperature structure of the model, we clearly have to worry 
about this problem.

\subsection{Abundance comparison of this work with EAGLNT}
A comparison in abundances between this work and EAGLNT for the 25 stars in common provides an
independent estimate of the errors of the derived abundances. The
results are summarized in Table~\ref{tb:abucom}.

\begin{table}
\caption{Mean abundance differences (this work$-$EAGLNT) and standard deviations.
N is the number of stars, for which a comparison was possible}
\label{tb:abucom}
\begin{tabular}{lrrr}
\noalign{\smallskip}
\hline
\noalign{\smallskip}
     & $<\Delta>$ & $\sigma$ & N \\
\noalign{\smallskip}
\hline
$\feh_I$ &$-$0.020  & 0.068 & 25 \\
$\feh_{II}$&$-$0.004& 0.090 & 25\\
$\OFe$&     0.147& 0.064 &  5\\
$\NaFe$&    $-$0.079& 0.057 & 21\\
$\MgFe$&    $-$0.020& 0.080 & 21\\
$\AlFe$&    $-$0.033& 0.080 & 16\\
$\SiFe$&    $-$0.012& 0.054 & 23\\
$\CaFe$&     0.055& 0.039 & 25\\
$\TiFe$&    $-$0.093& 0.099 & 24\\
$\NiFe$&    $-$0.008& 0.045 & 25\\
$\BaFe$&     0.068& 0.081 & 25\\
\noalign{\smallskip}
\hline
\end{tabular}
\end{table}

The agreement in iron abundance derived from \Feone\ lines is
satisfactory with deviations within $\pm$0.1 dex for the 25
common stars. These small deviations are mainly explained by
different temperatures 
given the fact that the abundance
differences increase with temperature deviations between the two
works. The rms deviation in iron abundance derived from \Fetwo\ lines are
slightly larger than that from \Feone\ lines. The
usage of different gravities partly explain this. But
the small line-to-line scatter from 8 \Fetwo\ lines in our work
indicates  a more reliable
abundance than that of EAGLNT who used 2 \Fetwo\ lines only.

Our oxygen abundances are systematically higher by 0.15~dex than those of
EAGLNT for 5 common stars.
Clearly, the temperature deviation is the main reason. The systematically
lower value of 70~K in our work increases [O/Fe] by 
0.10 dex (see Table~\ref{tb:abuerr}).

The mean abundance differences for Mg, Al, Si, Ca and
Ni between the two works are hardly significant.
The systematical differences (this work -- EAGLNT) of $-$0.08 dex for
[Na/Fe] and [Ti/Fe] and $+$0.07~dex for [Ba/Fe] are difficult to
explain, but we note that when the abundances are based on a few
lines only, a systematic offset of the stars relative to the Sun
may occur simply because of errors in the solar equivalent
widths. 

\section{Stellar masses, ages and kinematics}
\label{sec:age}

\subsection{Masses and ages}
\begin{figure}
\resizebox{\hsize}{!}{\includegraphics{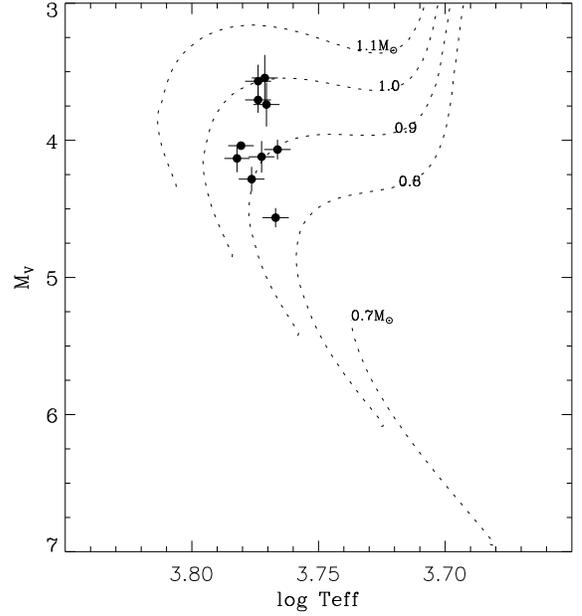}}
\caption{The positions of our program stars with $-0.77<\feh<-0.66 $ compared to the evolutionary tracks
of VandenBerg et al. (\cite{VandenBerg99}) with $\feh = -0.71$.}
\label{fig:evol97m76}
\end{figure}

As described in Sect.~\ref{subsec:logg}, the stellar mass is required
in the determination of the gravity from the Hipparcos
parallax. With the derived temperature
and absolute magnitude, the mass was estimated from the stellar
position in the $ M_{V} - \logteff$ diagram (see Fig.~\ref{fig:evol97m76}) 
by interpolating in
the evolutionary tracks of VandenBerg et al. (\cite{VandenBerg99}), which are
distributed in metallicity with a step of $\sim 0.1$ dex. These new tracks
are based on the recent OPAL opacities (Rogers \& Iglesias \cite{Rogers92})
using a varying helium abundance with [$\alpha$/Fe] = 0.30
for $\feh \leq -0.3 $ and a constant helium abundance ($ Y=0.2715 $)
without $\alpha$ element enhancement for
$\feh \geq -0.2$. 
Fig.~\ref{fig:evol97m76} shows the position of our
program stars with $-0.77<\feh<-0.66 $ compared to the evolutionary
tracks of Z = 0.004 ($ \feh =-0.71$). The errors in \teff , $M_V$, and \feh\
translate to an error of $0.06 M_{\odot}$ in the mass.
 
\begin{figure}
\resizebox{\hsize}{!}{\includegraphics{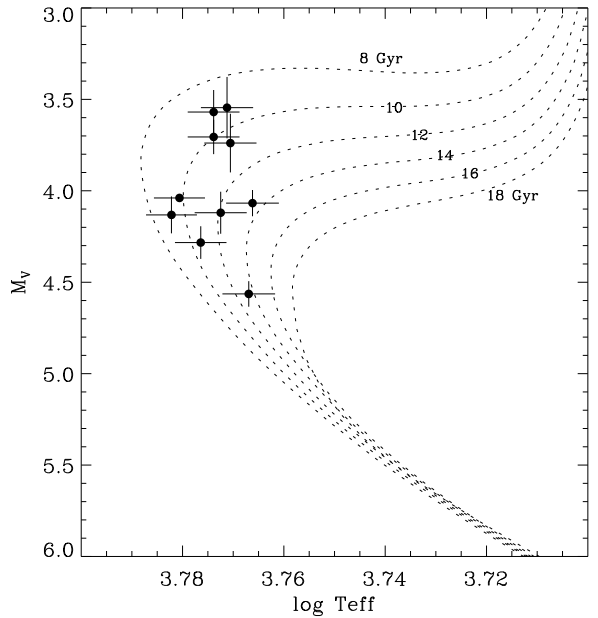}}
\caption{The positions of our program stars with $-0.77<\feh<-0.66 $ compared to the isochrones
of VandenBerg et al. (\cite{VandenBerg99}) with $\feh = -0.71$.}
\label{fig:iso97m76}
\end{figure}

Stellar age is an important parameter when studying
the chemical evolution of the Galaxy as a function of
time. Specifically, the age is useful in order to
interpret abundance ratios as a function of metallicity.
In this work, the stellar age was obtained simultaneously with the mass
from interpolation in the evolutionary tracks of
VandenBerg et al. (\cite{VandenBerg99}).
It was checked that practically the same age is
derived from the corresponding isochrones. As an example, a set of stars
are compared to isochrones in Fig.~\ref{fig:iso97m76}.
The error of the age due to the uncertainties of
\teff , $M_V$, and \feh\ is about
15\% ($\sigma(\log \tau)$ = 0.07) except for a few stars, which have 
relatively large errors of the Hipparcos parallaxes.
  
\subsection{Kinematics}

   Stars presently near the Sun may come from a wide range of 
Galactic locations. Information on their origin  will help us to 
understand their abundance ratios. Therefore, stellar space
velocity, as a clue to the origin of a star in the Galaxy, is very
interesting.

The accurate distance and proper motion
available in the Hipparcos Catalogue (ESA \cite{ESA97}), combined
with stellar radial velocity, make it
possible to derive a reliable space velocity. 
Radial velocities from the CORAVEL survey for 53 stars were kindly made available
by Nordstr\"{o}m (Copenhagen) before
publication. 
These velocities are compared with our values derived from the Doppler shift of
spectral lines.
A linear least squares fit for 40 stars (excluding the suspected binaries)
gives:
\begin{eqnarray}
RV =0.997\left(\pm0.002\right)RV_{\mathrm{CORAVEL}}+0.26\left(\pm0.12\right)
\kmprs  \nonumber
\end{eqnarray}
The rms scatter around the relation is 0.72 \kmprs, showing
that our radial velocities
are as accurate as 0.5 \kmprs. Hence, our values are adopted
for stars not included in the CORAVEL survey.

The calculation of the space velocity with respect to the Sun
is based on the method 
presented by Johnson \& Soderblom (\cite{Johnson87}).
The correction of space velocity to the Local Standard of
Rest is based on a solar motion,
($-$10.0,$+$5.2,$+$7.2) \kmprs\ in (U,V,W)\footnote{In the present work,
U is defined to be positive in the anticentric direction.}, as derived from Hipparcos
data by Dehnen \& Binney (\cite{Dehnen98}).
The error in the space velocity arising from the uncertainties 
of distance, proper motion
and radial velocity is very small with a value of
about $\pm 1$ \kmprs .

The ages and space velocities derived in the present work
are generally consistent with EAGLNT. But the more accurate
absolute magnitude, as well as the
new set of theoretical isochrones, in our study should give more
reliable ages than those determined by EAGLNT based on the photometric
absolute magnitude and the old isochrones of
VandenBerg \& Bell (\cite{VandenBerg85}). This situation is also true for
space velocities with our results based on distances  and proper motions
now available from Hipparcos.

\section{Results and discussion}
\label{sec:result}
\subsection{Relations between abundances, kinematics and ages}
The observed trends between  abundances, kinematics and ages are
the most important information for theories of
Galactic evolution. Especially, EAGLNT have provided many new
results on this issue. For example, the substantial dispersion in
the AMR 
found by EAGLNT argues against the
assumption of chemical homogeneity adopted in many chemical
evolution models. It is, however, important to test the results
of EAGLNT for a different sample of disk stars. Based on
more reliable ages and kinematics, the present study makes such an
investigation.

\subsubsection{Age-metallicity relation in the disk}

\begin{figure}
\resizebox{\hsize}{!}{\includegraphics{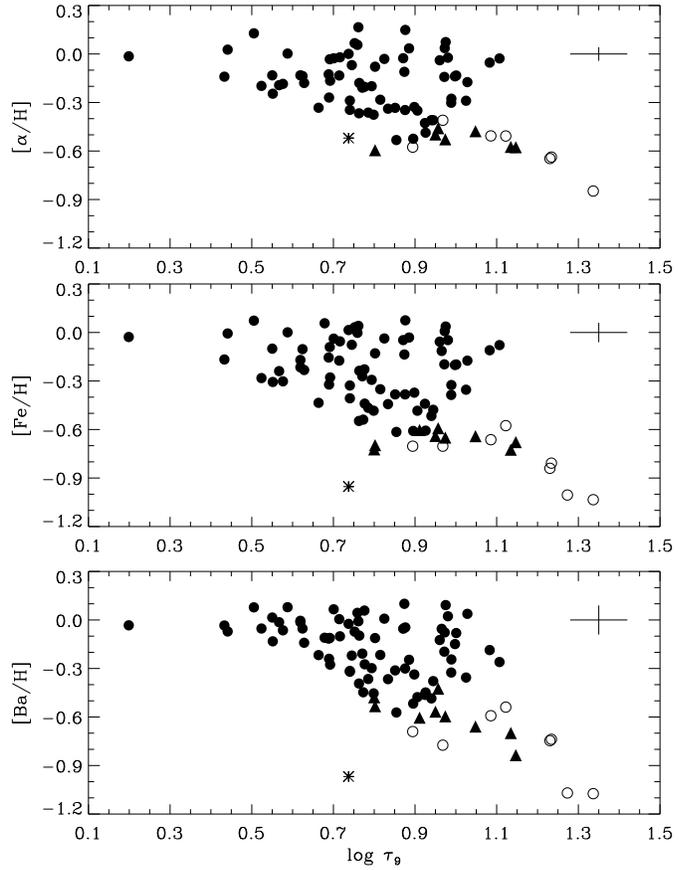}}
\caption{The abundances of $\alpha$ elements, Fe and Ba as a function of
  logarithmic age. The symbols are shown in
  Fig.~\ref{fig:VlsrFe}.}
\label{fig:AbuAge}
\end{figure}

Fig.~\ref{fig:AbuAge} 
shows the age-metallicity relations for $\alpha$, iron\footnote{Here and in
the following sections and figures, the
iron abundance is the mean abundance derived from
all \Feone\ and \Fetwo\ lines with equal weight to each line.} and barium elements, where $\alpha$ represents the mean abundance of Mg, Si, Ca and Ti. Generally,
there is a loose correlation between age and abundance. Stars younger than 5 Gyr ($\log \tau_{9} < 0.7$) are
more metal-rich than $\feh \simeq -0.3$, and stars with $\feh < -0.5$ 
are not younger than 6-7 Gyr ($\log \tau_{9} > 0.8$). The
deviating young halo star
\object{HD\,97916} (indicated by an asterisk in
Fig.~\ref{fig:AbuAge}) is discussed in Sect.~\ref{sec:conclusion}.

The correlation between age and abundance is, however, seriously distorted by 
a considerable scatter. Stars with
solar metallicity have an age spread as large as 10 Gyr, and
coeval stars at 10 Gyr show metallicity differences as high as 0.8 dex.
Such a dispersion cannot be explained by either the abundance
error ($< 0.1$ dex) or the age uncertainty ($\sim 15$\%) in the AMR. 
This is an important constraint on GCE models, which must reproduce
both the weak correlation and the substantial dispersion.

It is seen from Fig.~\ref{fig:AbuAge} that Ba has the steepest slope in
the AMR, Fe has intermediate slope, and the $\alpha$ elements show only
a very weak trend with \feh . This was also found by EAGLNT and
is consistent with nucleosynthesis theory that suggests
that the main synthesis sites of Ba, Fe and  $\alpha$
elements are AGB stars (1-3 $M_{\odot} $), SNe Ia (6-8
$M_{\odot}$) and SNe II ($> 8M_{\odot} $), respectively.
Due to their longer lifetime, lower mass stars contribute to the enrichment
of the Galaxy at a later epoch, i.e. after massive stars have been
polluting their products into the ISM. Hence, the Ba abundance is relatively low in
the beginning of the disk evolution and increases quickly in the
late stage, leading to a steeper slope.

It is interesting that there is a hint of a smaller metallicity spread for young stars
with $\log \tau_{9}=0.4-0.8$ in this work than in EAGLNT,
while the spread is similar for old stars.
If we are not misled by our sample (less young stars than
in the
EAGLNT sample and a lack of stars with $\log \tau_{9}<0.4$), it
seems that there are metal-rich stars at any time in the solar
neighbourhood while metal-poor stars are always old.
Another interesting feature for the young stars is that [Ba/H]
has a smaller metallicity spread than [Fe/H] and
[$\alpha$/H]. This could be due to the dependence of elemental yield on
the progenitor's mass. Ba is produced by AGB stars
with a rather small mass range of 1-3 $M_{\odot}$, while Fe and $\alpha$
elements are
synthesized by SNe having a mass range $\sim$6-30 $M_{\odot}$.

\subsubsection{Stellar kinematics as functions of age and metallicity}

The study of the dispersion in kinematical parameters as a function of Galactic
time is more interesting than the kinematical data
alone, because any abrupt increase in dispersion may indicate special
Galactic processes occurring during the evolution. Generally, dispersions in $\Vlsr$,
$\Wlsr$ and total velocity
increase with stellar age. We have not enough stars 
at $\sim$ 2.5 Gyr and 10 Gyr to confirm the abrupt increases in the $\Wlsr$ 
dispersion found by EAGLNT at these ages. 
Instead, our data
seems to indicate that the kinematical dispersion (possibly also
the metallicity) is fairly
constant for stars younger than 5 Gyr ($\log \tau_{9}=0.7$), but
it increases with 
age for stars with $\log \tau_{9}>0.7$. Coincidentally, 5 Gyr corresponds
to $\feh \simeq -0.4$ dex, the
metallicity  where EAGLNT suggested
an abundance transition related to a dual formation of the Galactic
disk. The abundance transition at $\feh \simeq -0.4$ dex is confirmed
by our data, but the increase of the $\Wlsr$ dispersion at
$\feh \simeq -0.4$ found by EAGLNT is less obvious in our data.

\begin{figure}
\resizebox{\hsize}{!}{\includegraphics{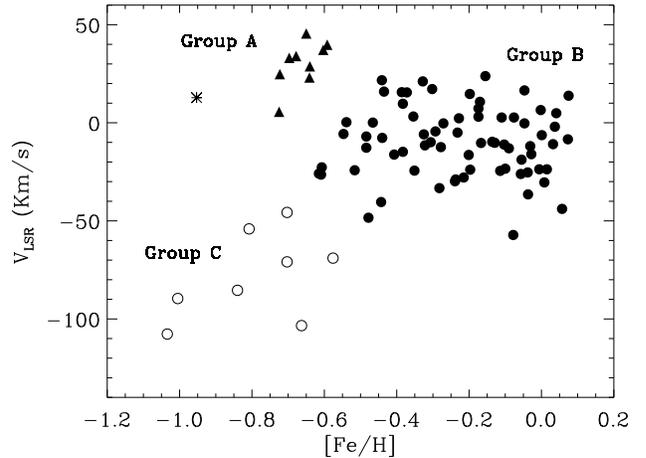}}
\caption{$\Vlsr$ vs. $\feh$ with different symbols showing three groups 
of stars. The asterisk indicates
the halo star HD\,97916 discussed in Sect.~\ref{sec:conclusion}.}
\label{fig:VlsrFe}
\end{figure}

 When the velocity component in the direction of Galactic rotation, $\Vlsr$, is investigated as a 
function of the metallicity (see Fig.~\ref{fig:VlsrFe}),
we find that there are two subpopulations for
$\feh \leq -0.6$ with positive $\Vlsr$ in 
group A and negative $\Vlsr$ in group C,
while stars with $ \feh \geq -0.6 $ have
$\Vlsr$ around $\Vlsr= -10 \kmprs$ (group B). The pattern persists when other
elements are substituted for Fe.
As shown in Edvardsson et al. (\cite{Edvardsson93b}), there is a
tight correlation between $\Vlsr$ and 
the mean Galactocentric distance in the stellar orbit, 
$R_{m}$. Hence, we can trace the metallicity at different
Galactocentric distances assuming that
$R_{m}$ is a reasonable estimator of the radius of the star's original
orbit. Note, however, that the lower metallicity toward
the Galactic center ($\Vlsr \lesssim -50 \kmprs $) for group C stars may be due to their large
ages. Excluding these stars, a trend of decreasing metallicity with 
increasing $\Vlsr$ for stars with similar age is found, which indicates a radial abundance gradient in the
disk, and thus suggests a faster evolution in the inner disk than
the outer. This is compatible with a higher
SFR, due to the higher density, in the inner disk.

There are two possibilities to explain the stars in group C. 
One is anchored to the fact that the oldest stars ($>$ 10 Gyr) 
in our sample have the lowest
$\Vlsr$, i.e. the smallest $R_{m}$, indicating that the Galaxy
did not extend to the Sun at 10 Gyr ago according to an inside-out
formation process of the Galaxy. 
The other is that these stars come from the thick disk, which is
older and more metal-poor than the thin disk.

\subsection{Relative abundances}
\begin{figure*}
\resizebox{\hsize}{!}{\includegraphics{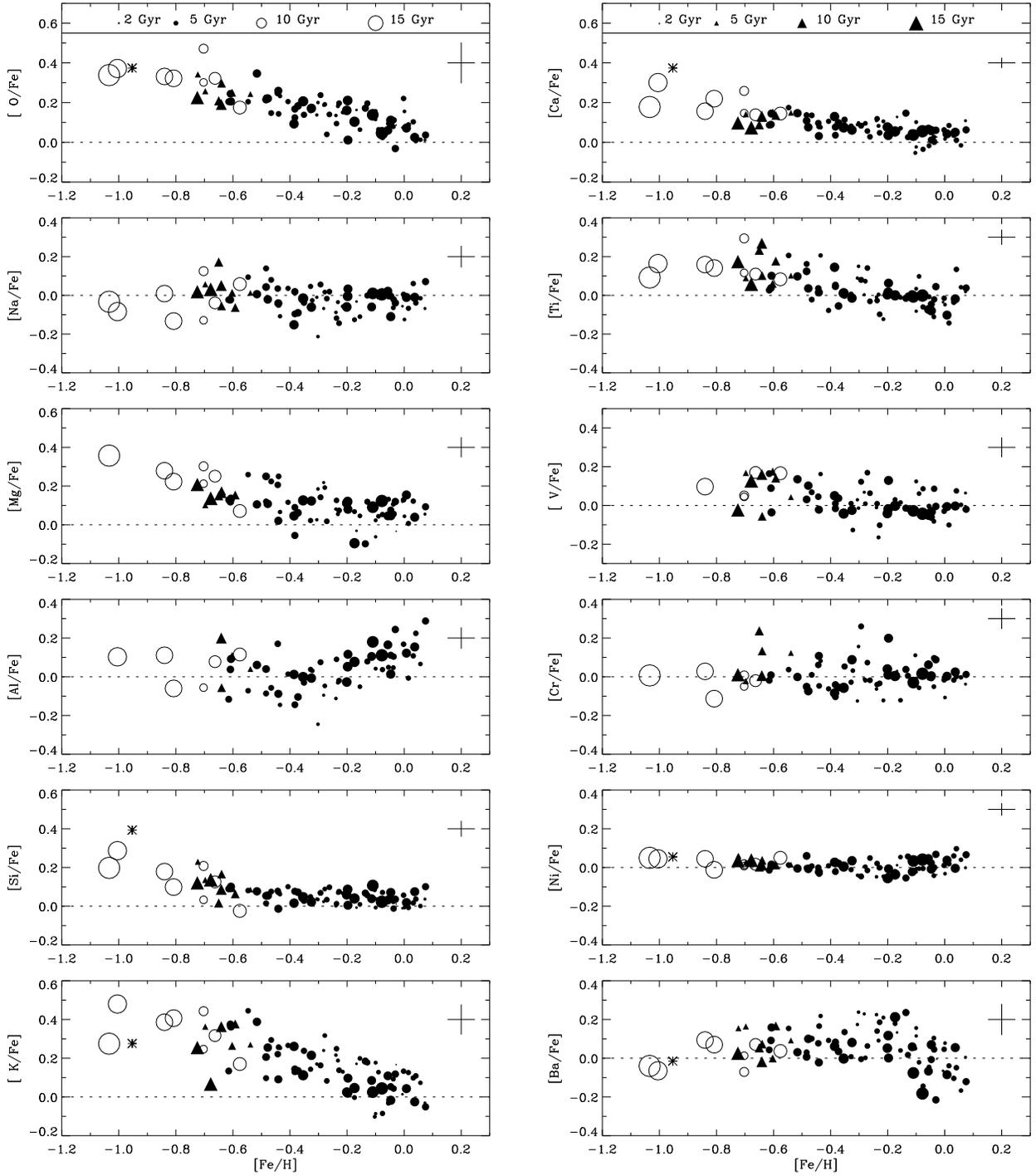}}
\caption{Abundance patterns for elements from O to Ba. The
symbols are the same as in Fig.~\ref{fig:VlsrFe} and their 
size is proportional to stellar age. Note that the trends of [Al/Fe] and [K/Fe]
may be spurious due to the neglect of non-LTE effects.}
\label{fig:AbuFea}
\end{figure*}

The general trends of elemental abundance with respect to iron 
as a function of metallicity, age and kinematics are to be
studied in connection with Galactic evolution models and
nucleosynthesis theory. The main results are shown in
Fig.~\ref{fig:AbuFea} and will be discussed together with those
of EAGLNT.

\subsubsection{Oxygen and magnesium}

In agreement with most works, [O/Fe] shows a tendency to decrease constantly 
with increasing metallicity for disk stars. As oxygen is only produced in the massive
progenitors of SNe II, Ib and Ic, it is mainly build up at early 
times of the Galaxy, leading to an overabundance of oxygen 
in halo stars. The [O/Fe] ratio gradually decline in the disk stars when iron
is produced by the long-lived SNe Ia. The time delay
of SNe Ia relative to SNe II is responsible for the continuous
decrease of oxygen in disk stars.
The tendency for [O/Fe] to continue to decrease at
$\feh > -0.3$ argues for an increasing ratio of SNe Ia
to SNe II also at the later stages of the disk evolution.

In general, the relation of [O/Fe] vs.
$\Vlsr$ reflects the variation of [Fe/H] with  $\Vlsr$ (see
Fig.~\ref{fig:VlsrFe}). [O/Fe] decreases with 
increasing $\Vlsr$ for stars with $\Vlsr < 0$ and slowly increase with further
larger $\Vlsr$. Considering their
similar ages, the decreasing [O/Fe] from group A to group B stars
may be attributed to the increasing $\Vlsr$, whereas the higher [O/Fe] of
group C is due to an older age.

The magnesium abundance shows a decreasing trend with 
increasing metallicity like oxygen for $\feh < -0.3$ but it tends to
flatten out for higher metalicities.
Given that magnesium is theoretically predicted to be formed only in
SNe II, the similar decreasing trend as oxygen is easily understood,
but the flat [Mg/Fe] towards higher
metallicities than $\feh > -0.3$ is unexpected.
It seems
that SNe II are not the only source for Mg. Perhaps SNe Ia also contribute 
to the enrichment of Mg during disk evolution. 

The flat trend of [Mg/Fe] vs. [Fe/H] for $\feh > -0.3$ is
also evident from the data of EAGLNT if the high Mg/Fe ratios
of their NaMgAl stars are reduced to a solar ratio as found by Tomkin
et al. (\cite{Tomkin97}). Feltzing and Gustafsson
(\cite{Feltzing99}) also find [Mg/Fe] to be independent of
metallicity for their more metal-rich stars although the scatter is large.

With more magnesium lines in the present study, we get a similar scatter
of [Mg/Fe] as EAGLNT. The scatter is slightly larger than that of
oxygen and in particular much larger than those of
Si and Ca. Although we do not find a large line-to-line
scatter in the Mg abundance determination, it is still unclear if
the scatter in [Mg/Fe] is cosmic. Only
3 \Mgone\ lines are available for most stars while Si and Ca are
represented by 20-30
lines. There is no obvious evidence showing the scatter to be an effect
of different $ \Vlsr$. Nor do we find a clear separation
of thick disk stars from thin disk stars in the diagram of [Mg/Fe] vs.
[Mg/H], as has been found by Fuhrmann (\cite{Fuhrmann98}).
It seems that neither observation nor theory is
satisfactory for Mg.

\subsubsection{Silicon, calcium and titanium}
Like magnesium, [Si/Fe] and [Ca/Fe] decrease with increasing metallicity for $\feh < -0.4$ and
then flatten out with further increasing [Fe/H]. The result is in
agreement with EAGLNT, who found a ``kink'' at $\feh=
-0.3\sim -0.2$. But [Ca/Fe] possibly continues to decrease for
$\feh > -0.4$ based on our data. The suspicion that Si is
about 0.05 dex overabundant relative to Ca for $\feh > -0.2$
and the possible upturn of silicon at higher metallicity in
EAGLNT are not supported by our data. 

Both Si and Ca have a very small star-to-star scatter (0.03~dex) at a given
metallicity for thin disk stars. The scatter is
slightly larger among the thick disk stars.
Since the scatter corresponds
to the expected error from the analysis, we conclude that
the Galactic scatter for [Si/Fe] and [Ca/Fe] is less than 0.03 dex in the
thin disk.

[Ti/Fe] was shown by EAGLNT to be a slowly
decreasing function of [Fe/H] and the decrease continues to higher
metallicity. Our data show a similar
trend but the continuous decrease toward higher metallicity is
less obvious with a comparatively large star-to-star scatter. There is no
evidence that the scatter is correlated with
$\Vlsr$.
We note that Feltzing \& Gustafsson (\cite{Feltzing99}) find
a similar scatter in [Ti/Fe] for metal-rich stars with the Ti
abundance based on 10-12 \Tione\ lines.

\subsubsection{Sodium and aluminum}

Na and Al are generally thought to be products
of Ne and C burning in massive stars.
The synthesis is
controlled by the neutron flux which in turn depends on the
initial metallicity and primarily on the initial O abundance.
Therefore, one expects a rapid increase of [Na/Mg] and [Al/Mg] with
metallicity. But our data shows that both Na and Al are poorly
correlated with Mg in agreement with EAGLNT.
This means that the odd-even effect has been greatly reduced in the
nucleosynthesis processes during the disk formation.

When iron is taken as the reference element, we find that [Na/Fe]
and [Al/Fe] are close to zero for $\feh < -0.2$, while EAGLNT
found 0.1-0.2 dex differences between $\feh
= -0.2$ and $\feh = -1.0$. Our results support the old
data by Wallerstein (\cite{Wallerstein62}) and Tomkin et al. (\cite{Tomkin85}), who suggested [Na/Fe] $\sim$ 0.0 for the
whole metallicity range of the disk stars.
The situation is the same for Al; EAGLNT found an overabundance of
[Al/Fe]~$\simeq 0.2$ for $\feh < -0.5$, whereas we find a solar ratio
for the low metallicity stars.
As discussed in Sect.~\ref{subsec:NLTE} this may, however, be due to a non-LTE
effect.

In the case of the more metal rich stars the abundance
results for Na and Al are rather confusing.
EAGLNT found that some metal-rich stars in the
solar neighbourhood are rich in Na, Mg and Al, but
the existence of such NaMgAl stars was rejected
by Tomkin et al. (\cite{Tomkin97}). Several further studies, however,
confirmed the
overabundance of some elements again. Porte de Morte
(\cite{Porte96}) found an overabundance of Mg but not of Na.
Feltzing \& Gustafsson (\cite{Feltzing99})
confirmed the upturn of [Na/Fe] but their metal-rich stars did
not show Mg and Al overabundances. In the present work we find a 
solar ratio of Na/Fe up to $\feh \simeq 0.1$, and a rather steep
upturn of [Al/Fe] beginning at $\feh \simeq -0.2$. As discussed
in Sect.~\ref{subsec:NLTE}, our Al abundances may, however, be severely affected
by non-LTE effects. We conclude that more accurate data on Na and
Al abundances are needed.

\subsubsection{Potassium}
  \KFe\ shows a decreasing trend with increasing metallicity for disk
stars. The result supports the previous work by Gratton \&
Sneden (\cite{Gratton87}) but our data have a smaller scatter.
Assuming that potassium is a product of explosive oxygen burning 
in massive stars, Samland (\cite{Samland98}) reproduces
the observed trend rather well.  Timmes et al. (\cite{Timmes95}),
on the other hand, predicts [K/Fe]$< 0.0$ for $\feh < -0.6$
in sharp contrast to the observations. Given that the
\Kone\ resonance line at $\lambda$7699, which are used to derive the K
abundances, is affected by non-LTE as discussed in Sect.~\ref{subsec:NLTE},
it seems premature to attribute K
to one of $\alpha$ elements.

\subsubsection{Vanadium, chromium and nickel}
V and Cr seem to follow Fe for the whole metallicity range with
some star-to-star scatter. The scatter is 
not a result of mixing stars with different $\Vlsr$, and the few very weak
lines used to determine the abundances prevent us to investigate
the detailed dependence on metallicity and to decide
if the scatter is cosmic or due to errors.

Ni follows iron quite well at all metallicities with 
a star-to-star scatter less than 0.03 dex.
Two features may be found after
careful inspection. Firstly, there is a hint that [Ni/Fe]
slightly decreases with increasing metallicity for  $-1.0 < \feh <
-0.2$. The trend is more clear, due to smaller star-to-star scatter,
than  in EAGLNT. Secondly, there is a subtle increase of [Ni/Fe]
for $ \feh > -0.2$.  Interestingly, Feltzing
\& Gustafsson (\cite{Feltzing99}) found a slight increase of [Ni/Fe]
towards even more metal-rich stars.

\subsubsection{Barium}
The abundance pattern of Ba is very similar to that of EAGLNT except for
a systematic shift of about +0.07 dex in [Ba/Fe].
Both works indicate a complicated dependence of
[Ba/Fe] on metallicity. First, [Ba/Fe] seems to increase slightly
with metallicity for $\feh < -0.7$, and
then keeps a constant small overabundance until $\feh \sim -0.2$, after which
[Ba/Fe] decreases towards higher metallicities.

Barium is thought to be synthesized by neutron capture
s-process in low mass AGB stars with an evolutionary timescale
longer than that of iron-producing SNe Ia. Therefore, [Ba/Fe]
is still slightly underabundant at $\feh = -1.0$.  Ba 
is then enriched significantly at later stages of the disk
evolution, but the decrease of [Ba/Fe] for more metal-rich stars
beginning with $\feh \sim -0.2$ is unexpected.

\begin{figure}
\resizebox{\hsize}{!}{\includegraphics{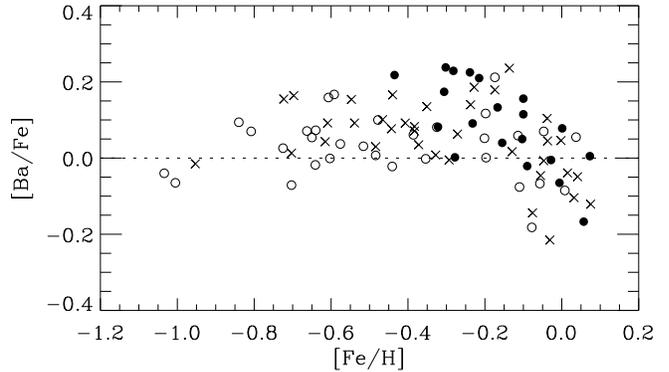}}
\caption{Different relations of [Ba/Fe] vs. [Fe/H] for stars with different age ranges: 
$\log \tau_{9} < 0.7$ (filled circles),
$\log \tau_{9} = 0.7 -0.9 $ (crosses) and $\log \tau_{9} > 0.9$ (open circles).}
\label{fig:BaAge}
\end{figure}

Given that the low [Ba/Fe] for some stars may  be related
to their ages, the relation of [Ba/Fe] vs. [Fe/H] at different
age ranges was investigated
(see Fig.~\ref{fig:BaAge}). In agreement with EAGLNT, the run of
[Ba/Fe] vs. [Fe/H] in old stars with $\log \tau_{9} > 0.9$
($\sim$8 Gyr) and  $0.7 <\log \tau_{9} < 0.9$ shows a flat
distribution for $\feh < -0.3$ and a negative slope for $\feh > -0.3$.
All young stars with $\log \tau_{9} < 0.7$ ($\sim$5 Gyr) have $\feh > -0.3$
and a clear decreasing trend of [Ba/Fe] with [Fe/H] is seen. In
addition, there is a hint of higher [Ba/Fe] for younger
stars both in the interval $-0.7 <\feh
<-0.3$, where [Ba/Fe] is constant and in the interval $\feh > -0.3$,
where [Ba/Fe] is decreasing.
This is consistent with the formation of young stars at a later
stage of the disk when long-lived AGB stars have enhanced
Ba in the ISM. The flat [Ba/Fe] for $\feh <-0.3$ may be
explained by the suggestion of EAGLNT that
the synthesis of Ba in AGB stars is independent of metallicity, i.e.
that Ba shows a primary behaviour during the evolution of
the disk. But the age effect alone cannot explain the
underabundant [Ba/Fe] in metal-rich stars, because [Ba/Fe]
decreases with metallicity for all ages after $\feh = -0.3$.
One reason could be that s-element synthesis
occurs less frequently in metal-rich AGB stars possibly because the
high mass loss finishes their evolution earlier. 

\section{Concluding remarks}
\label{sec:conclusion}
One of the interesting results of this study is that the
oldest stars presently located in the solar neighbourhood have
$\Vlsr \lesssim -50 \kmprs$. Hence, they probably originate from the inner disk
having $\Rm < 7$ kpc.
This is not coincidentally found in our study. The EAGLNT sample contains about
20 such stars. As shown in both works, these stars are generally
more metal-poor than other stars and they show a larger spread in
[Fe/H] and [Ba/H] than in [$\alpha$/H] (see Fig.~\ref{fig:AbuAge}).
According to EAGLNT, they have
higher [$\alpha$/Fe] than other disk stars at a
metallicity about $\feh=-0.7$. 

Considering these different properties, we suggest that they
do not belong to the thin disk. Firstly, they are older (10-18
Gyr) than other stars.  Secondly, if they are thin disk
stars, it is hard to understand why stars coming
from both sides of the solar annulus have lower metallicity than
the local region. Thirdly, these stars show a relatively
small metallicity dispersion
at such early Galactic time, i.e. smaller than stars at 8-10 Gyr.
This is not in agreement with the effect of orbital diffusion working
during the evolution of the thin disk, which
suggests larger metallicity dispersion for older stars. Finally,
the $\Wlsr$ dispersion of these stars is about 40~$\kmprs$,
considerably larger than the typical value of about 20~$\kmprs$ 
for thin disk stars. 
Consistently, the kinematics, age, metallicity and abundance
ratios of these stars follow the features of the thick 
disk:  $\Vlsr \lesssim -50 \kmprs$, $\sigma(\Wlsr) \simeq 40 \kmprs$,
$\tau > 10$ Gyr, $\feh < -0.5$ and [$\alpha$/Fe] $\sim 0.2$. We
conclude that these oldest stars in both
EAGLNT and this work are thick disk stars. Hence, they are probably
not resulting from an inside-out formation of the
Galactic disk, but have been formed in connection with
a merger of satellite components with the Galaxy. 

\begin{figure}
\resizebox{\hsize}{!}{\includegraphics{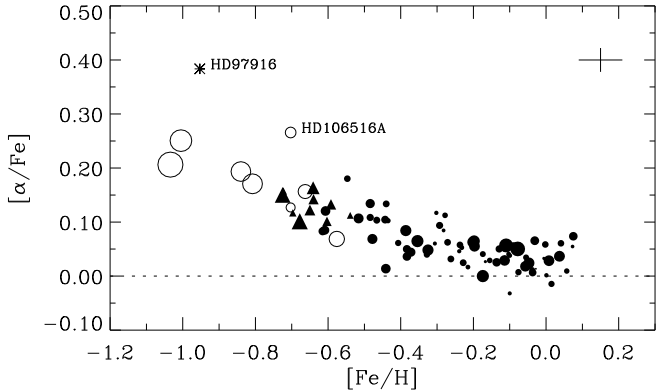}}
\caption{The mean $\alpha$ (Mg, Si, Ca and Ti) abundance as a
function of metallicity. The symbols are the same as in Fig.~\ref{fig:VlsrFe} and
their size is proportional to stellar age.}
\label{fig:AbuaFea}
\end{figure}

Concerning the abundance connection of the thick disk with the thin disk, our
data for [$\alpha$/Fe], shown in Fig.~\ref{fig:AbuaFea}, suggest a more
smooth trend than those of EAGLNT, who found a correlation between
[$\alpha$/Fe] and $\Rm$ at $\feh
\sim -0.7$. We leave the issue open considering the small
number of these stars in our work. Two stars marked by their names in
Fig.~\ref{fig:AbuaFea} may be particularly interesting because they
show significantly higher [$\alpha$/Fe] than other stars.
Fuhrmann \& Bernkopf (\cite{Fuhrmann99})
suggest that one of them, \object{HD\,106516A}, is a thick-disk
field blue straggler. It is unclear if this can explain
the higher [$\alpha$/Fe].
\object{HD\,97916} is a nitrogen rich binary (Beveridge \& Sneden
\cite{Beveridge94}) with $\Ulsr=-117 \kmprs$ and $\Wlsr=101 \kmprs$
(typical for halo stars), but with $\Vlsr=22 \kmprs$ 
similar to the value for thin disk stars. Surprisingly,
this star is also very young (5.5 Gyr) for it's metallicity. 

It is interesting to re-inspect the observational results
for thin disk stars excluding the thick disk stars.
More direct information on
the evolution of the Galactic thin disk will be then obtained. In summary,
the thin disk
is younger (not older than 12 Gyr), more metal-rich
($\feh>-0.8$) and has a smaller [$\alpha$/Fe] spread (0.1 dex)
without the mixture of the thick disk stars. In
particular, the AMR is more weak and there seems to exist a radial
metallicity gradient. All these features agree better with the present
evolutionary models for the Galactic disk. 

We emphasize here that there is no obvious 
gradient in [$\alpha$/Fe] for
the thin disk at a given metallicity. Such a gradient was
suggested by EAGLNT based on higher [$\alpha$/Fe] of the
oldest stars with $\Rm < 7$ kpc than stars  with $\Rm > 7$
kpc (see their Fig. 21). After we have ascribed these oldest
stars to the thick disk, the abundance gradient disappears.

Our study of relative abundance ratios as a function of $\feh$
suggests that there are subtle differences of origin and enrichment history
both within the group of $\alpha$ elements and the iron-peak elements. 
Nucleosynthesis theory predicts that Si and
Ca are partly synthesized in SNe Ia, while O and Mg are only
produced in SNe II (Tsujimoto et al. \cite{Tsujimoto95}).
Our data suggest, however,
that SNe Ia may also be a significant synthesis site of Mg, because [Mg/Fe]
shows a trend more similar to [Si/Fe] and [Ca/Fe] than to [O/Fe].
Ti may not lie in
a smooth extension of Si and Ca, because there is a hint of a
decrease of [Ti/Fe] for $\feh > -0.4$ not seen in the case of Si and Ca.
The situation for the odd-Z elements is more complicated.
The available data for Na and Al show confusing disagreements; EAGLNT
finds an overabundance of 0.1 to 0.2 dex for [Na/Fe] and [Al/Fe] 
among the metal-poor disk stars, whereas our study points at solar
ratios. Two other odd-Z elements, K and Sc (Nissen et al. \cite{Nissen99}),
behave like $\alpha$ elements, but the result for K is sensitive to the
assumption of LTE.
The iron-peak elements also show different behaviours: V, Cr and Ni
follow Fe very well, while [Mn/Fe] (Nissen et al. \cite{Nissen99})
decreases with decreasing metallicity from [Mn/Fe]~$\simeq 0.0$ at
$\feh = 0.0$ to [Mn/Fe]~$\simeq -0.4$ at $\feh = -1.0$.
We conclude that the terms ``$\alpha$ elements'' and ``iron-peak
elements'' do not indicate productions in single processes,
and that each element seems to have a unique enrichment history.

\section*{Acknowledgements}
This research was supported by the Danish Research Academy and
the Chinese Academy of Sciences. Bengt Edvardsson is thanked for
providing a grid of the Uppsala new MARCS model atmospheres, and Birgitta
Nordstr\"{o}m for communicating CORAVEL radial velocities in advance
of publication.


\begin{thebibliography}{}
\bibitem[1995]{Alonso95}
Alonso A., Arribas S., Mart\'{\i}nez-Roger C., 1995, A\&A 297, 197
\bibitem[1996]{Alonso96}
Alonso A., Arribas S., Mart\'{\i}nez-Roger C., 1996, A\&A 313, 873
\bibitem[1995]{Anstee95}
Anstee S.D., O'Mara B.J., 1995, MNRAS 276, 859
\bibitem[1997]{Asplund97}
Asplund M., Gustafsson B., Kiselman D., Eriksson K., 1997, A\&A 318, 521
\bibitem[1999]{Asplund99}
Asplund M., Nordlund \AA., Trampedach R., Stein R.F., 1999, A\&A 346, L17
\bibitem[1994]{Bard94}
Bard A., Kock M., 1994, A\&A 282, 1014
\bibitem[1991]{Bard91}
Bard A., Kock A., Kock M.,  1991, A\&A 248, 315
\bibitem[1998]{Baumuller98}
Baum\"{u}ller D., Butler K., Gehren T., 1998, A\&A 338, 637
\bibitem[1997]{Baumuller97}
Baum\"{u}ller D., Gehren T., 1997, A\&A 325, 1088 
\bibitem[1994]{Beveridge94}
Beveridge C.R., Sneden C., 1994, AJ 108, 285
\bibitem[1991]{Biemond91}
Bi\'emont E., Baudoux M., Kurucz R.L., Ansbacher W., Pinnington E.H., 1991, A\&A 249, 539
\bibitem[1976]{Biehl76}
Biehl D., 1976, Diplomarbeit, Inst. f. Theor. Physik u. Sternwarte, Kiel University
\bibitem[1986a]{Blackwell86a}
Blackwell D.E., Booth A.J., Menon S.L.R., Petford A.D.,  1986a, MNRAS 220, 289
\bibitem[1986b]{Blackwell86b}
Blackwell D.E., Booth A.J., Menon S.L.R., Petford A.D.,  1986b, MNRAS 220, 303
\bibitem[1982a]{Blackwell82a}
Blackwell D.E., Menon S.L.R., Petford A.D., Shallis M.J., 1982a, MNRAS 201, 611
\bibitem[1982b]{Blackwell82b}
Blackwell D.E., Petford A.D., Shallis M.J., Simmons G.J., 1982b, MNRAS 199, 43 
\bibitem[1982c]{Blackwell82c}
Blackwell D.E., Petford A.D., Simmons G.J., 1982c, MNRAS 201, 595
\bibitem[1994]{Carney94}
Carney B.W., Latham D.W., Laird J.B., Aguilar L.A., 1994, AJ 107, 2240
\bibitem[1990]{Chang90}
Chang T.N., 1990, Phys. Rev. A41, 4922
\bibitem[1998]{Dehnen98}
Dehnen W., Binney J.J., 1998, MNRAS 298, 387
\bibitem[1997]{ESA97}
ESA, 1997, The Hipparcos and Tycho Catalogues, ESA SP-1200
\bibitem[1993a]{Edvardsson93a}
Edvardsson B., Andersen J., Gustafsson B., Lambert D.L., Nissen
P.E., Tomkin J., 1993a, A\&A 275, 101 (EAGLNT)
\bibitem[1993b]{Edvardsson93b}
Edvardsson B., Gustafsson B., Nissen P.E., Andersen J., Lambert
D.L., Tomkin, J., 1993b, in Panchromatic View of Galaxies,
eds. Hensler, G., Theis, Ch. \& Gallagher, J., p. 401
\bibitem[1962]{Eggen62}
Eggen O.J., Lynden-Bell D., Sandage A.R., 1962, ApJ 136, 748
\bibitem[1999]{Feltzing99}
Feltzing S., Gustafsson B., 1999, A\&AS 129, 237
\bibitem[1996]{Francois96}
Fran\c{c}ois P., 1996, A\&A 313, 229
\bibitem[1998]{Fuhrmann98}
Fuhrmann K., 1998, A\&A 338, 161
\bibitem[1999]{Fuhrmann99}
Fuhrmann K., Bernkopf J., 1999, A\&A 347, 897
\bibitem[1973]{Garz73}
Garz T., 1973,  A\&A, 26, 471
\bibitem[1987]{Gratton87}
Gratton R.G., Sneden C., 1987, A\&A 178, 179
\bibitem[1992]{Hannaford92}
Hannaford P., Lowe R.M., Grevesse N., Noels A., 1992, A\&A 259, 301 
\bibitem[1990]{Holweger90}
Holweger H., Heise C., Kock M., 1990, A\&A 232, 510
\bibitem[1999]{IAU99}
IAU 1999, in Transactions of the IAU, vol. XXIII B, ed. J.Andersen, p.141
\bibitem[1987]{Johnson87}
Johnson D.R.H.,  Soderblom D.R., 1987,  AJ 93,  864
\bibitem[1982]{Kostyk82}
Kostyk R.I., 1982, Astrometriya Astrofiz. 46, 58
\bibitem[1978]{Lambert78}
Lambert D.L., 1978, MNRAS 182, 249
\bibitem[1968]{Lambert68}
Lambert D.L., Warner B., 1968, MNRAS 138, 181
\bibitem[1975]{Mackle75}
M\"ackle R., Holweger H., Griffin R., Griffin R., 1975, A\&A 38, 239
\bibitem[1987]{Magain87}
Magain P., 1987, A\&A 181, 323
\bibitem[1999]{Mashonkina99}
Mashonkina L., Gehren T., Bikmaev I., 1999, A\&A 344, 221
\bibitem[1968]{Mihalas68}
Mihalas D., Routly P.M., 1968, Galactic Astronomy, (Freeman, San Francisco), p. 101
\bibitem[1966]{Moore66}
Moore C.E., Minnaert M.G.J., Houtgast J., 1966, The Solar Spectrum
2935 \AA\  to 8770 \AA, National Bureau of Standards, Monograph
61, Washington
\bibitem[1988]{Morris88}
Morris D.H., Mutel R.L., 1988, AJ 95, 204
\bibitem[1992]{Nissen92}
Nissen P.E., Edvardsson B., 1992, A\&A 261, 255
\bibitem[1997]{Nissen97}
Nissen  P.E., Schuster W.J., 1997, A\&A 326, 751
\bibitem[1999]{Nissen99}
Nissen P.E., Chen Y.Q., Schuster W.J., Zhao G., 1999, A\&A (in press)
\bibitem[1991]{OBrian91}
O'Brian T.R., Wickliffe M.E., Lawler J.E., Whaling W., Brault J.W. 1991, J. Opt. Soc. Am. B 8, 1185
\bibitem[1983]{Olsen83}
Olsen E.H., 1983, A\&AS 54, 55 
\bibitem[1988]{Olsen88}
Olsen E.H., 1988, A\&A 189, 173
\bibitem[1993]{Olsen93}
Olsen E.H., 1993, A\&A 102, 89 
\bibitem[1990]{Petit90}
Petit M., 1990, A\&AS 85, 971
\bibitem[1996]{Porte96}
Porte de Mello G.F., da Silva L., 1996, 
in: Stellar Abundance, eds. B. Barbuy, W.J. Maciel
J.C. Greg\'orie-Hetem J.C. p. 59
\bibitem[1992]{Rogers92}
Rogers F.J., Iglesias C.A., 1992, ApJS 79, 507
\bibitem[1999]{Reetz98}
Reetz J., 1999, in: Galaxy Evolution: Connecting the Distant Universe with the Local Fossil Record, eds. M. Spite, F. Crifo, Kluwer (in press) 
\bibitem[1998]{Ryan98}
Ryan S.G., 1998, A\&A 331, 1051
\bibitem[1998]{Samland98}
Samland M., 1998, ApJ 496, 155
\bibitem[1989]{Schuster89}
Schuster W.J.,  Nissen P.E., 1989,  A\&A 221,  65
\bibitem[1982]{Simmons82}
Simmons G.J., Blackwell D.E., 1982,  A\&A 112, 209
\bibitem[1981]{Smith81}
Smith G., Raggett D.St.J., 1981, J.Phys. B 14, 4015
\bibitem[1985]{Steffen85}
Steffen M., 1985, A\&AS 59, 403
\bibitem[1996]{Takeda96}
Takeda Y., Kato K.-I., Watanabe Y., Sadakane K., 1996, PASJ 48, 511
\bibitem[1999]{Thevenin99}
Th\'{e}venin F., Idiart T.P., 1999, ApJ 521, 753
\bibitem[1995]{Timmes95}
Timmes F.X., Woosley S.E., Weaver T.A. 1995, ApJS 98, 617
\bibitem[1997]{Tomkin97}
Tomkin J., Edvardsson B., Lambert D.L., Gustafsson B., 1997, A\&A 327, 587
\bibitem[1985]{Tomkin85}
Tomkin J., Lambert D.L., Balachandran S., 1985, ApJ 290, 289
\bibitem[1995]{Tsujimoto95}
Tsujimoto T., Nomoto K., Yoshii Y., Hashimoto M., Yanagida S.,
Thielemann F.-K. 1995, MNRAS 277, 945
\bibitem[1980]{Twarog80}
Twarog B.A., 1980, ApJ 242, 242
\bibitem[1955]{Unsold55}
Uns\"old A., 1955, Physik der Sternatmosph\"aren, Springer-Verlag, Berlin
\bibitem[1985]{VandenBerg85}
VandenBerg D.A, Bell R.A. 1985, ApJS 58, 561
\bibitem[1999]{VandenBerg99}
VandenBerg D.A., Swenson F.J., Rogers F.J., Iglesias C.A., Alexander D.R., 1999 (in preparation)
\bibitem[1962]{Wallerstein62}
Wallerstein G., 1962, ApJS 6, 407
\bibitem[1985]{Whaling85}
Whaling W., Hannaford P., Lowe R.M., Bi\'emont E., Grevesse N., 1985, A\&A 153, 109
\bibitem[1997]{Wickliffe97}
Wickliffe M.E., Lawler J.E., 1997, ApJS 110, 163
\bibitem[1980]{Wiese80}
Wiese W.L., Martin G.A., 1980, in: Wavelengths and Transition
Probabilities for Atoms and Atomic Ions, NSRDS-NBS 68,
Washington, DC 
\bibitem[1998]{Zhao98}
Zhao G., Butler K., Gehren T., 1998, A\&A 333, 219
\bibitem[1999]{Zhao99}
Zhao G., Gehren T., 1999, in: The Galactic Halo: from Globular
Clusters to Field Stars, the 35th Li\'{e}ge International
Astrophysics Colloquium, eds. A. Noels, P. Magain (in press)
\end{thebibliography}
\end{document}